\documentclass[11pt]{article}

\usepackage{graphicx}
\usepackage{subfigure}
\begin{document}
\oddsidemargin .03in
\evensidemargin 0 true pt
\topmargin -.4in


\def\ra{{\rightarrow}}
\def\a{{\alpha}}
\def\b{{\beta}}
\def\l{{\lambda}}
\def\eps{{\epsilon}}
\def\T{{\Theta}}
\def\t{{\theta}}
\def\co{{\cal O}}
\def\car{{\cal R}}
\def\caf{{\cal F}}
\def\cs{{\Theta_S}}
\def\pr{{\partial}}
\def\tri{{\triangle}}
\def\na{{\nabla }}
\def\S{{\Sigma}}
\def\s{{\sigma}}
\def\sp{\vspace{.15in}}
\def\hs{\hspace{.25in}}

\newcommand{\be}{\begin{equation}} \newcommand{\ee}{\end{equation}}
\newcommand{\bea}{\begin{eqnarray}}\newcommand{\eea}
{\end{eqnarray}}


\begin{titlepage}
\topmargin= -.2in
\textheight 9.5in

\begin{center}
\baselineskip= 18 truept
\vspace{.3in}

\centerline{\Large\bf Quantum Kerr(Newman) degenerate stringy vacua in ${\mathbf{4D}}$ on a non-BPS brane}

\vspace{.6in}
\noindent
{ {\bf Sunita Singh},{\bf K. Priyabrat Pandey}, {\bf Abhishek K. Singh} {\bf and} {{\bf Supriya Kar}\footnote{skkar@physics.du.ac.in }}}

\vspace{.2in}

\noindent

\noindent
{{\Large Department of Physics \& Astrophysics}\\
{\Large University of Delhi, New Delhi 110 007, India}}

\vspace{.2in}

{\today}
\thispagestyle{empty}

\vspace{.6in}
\begin{abstract}
We investigate some of the quantum gravity effects on a vacuum created pair of $(D{\bar D})_3$-brane by a non-linear $U(1)$ gauge theory on a $D_4$-brane. In particular we obtain a four dimensional quantum Kerr(Newman) black hole in an effective torsion curvature formalism sourced by 
a two form dynamics in the world-volume of a $D_4$-brane on $S^1$. Interestingly the event horizon is found to be independent of a non-linear electric charge and the $4D$ quantum black hole is shown to describe a degenerate vacua in string theory. We show that the quantum Kerr brane universe 
possesses its origin in a de Sitter vacuum. In a nearly $S_2$-symmetric limit the Kerr geometries may seen to describe a Schwarzschild and Reissner-Nordstrom quantum black holes. It is argued that a quantum Reissner-Nordstrom tunnels to a large class of degenerate Schwazschild vacua. 
In a low energy limit the non-linear electric charge becomes significant at the expense of the degeneracies. In the limit the quantum geometries 
may identify with the semi-classical black holes established in Einstein gravity. Analysis reveals that a quantum geometry on a vacuum created $D_3$-brane universe may be described by a low energy perturbative string vacuum in presence of a non-perturbative quantum correction.

\baselineskip=14 truept

\vspace{.12in}

\vspace{1in}

\noindent

\noindent

\end{abstract}
\end{center}

\vspace{.2in}

\baselineskip= 16 truept

\vspace{1in}

\end{titlepage}

\baselineskip= 18 truept

\section{Introduction}
Conceptual ideas to explore a theory of quantum gravity have been in the folklore of theoretical physics for quite some time. There are a number of
different techniques to investigate the quantum gravity effects to Einstein gravity \cite{hawking,gross,thooft,thooft-2}. It is believed that a quantum gravity may be described by a background independent non-perturbation theory. Interestingly an eleven dimensional M-theory has been conjectured \cite{witten-M}. It attempts to unify five distinct string theories in ten dimensions. In a low energy limit, M-theory reduces to a supergravity theory. Furthermore M-theory on $S^1$ reduces to a type IIA superstring theory, where the even Dirichlet ($D$) branes are dynamical objects. Importantly a gravity theory in bulk and a gauge theory on its boundary have been established as an insightful tool using a holographic correspondence \cite{susskind}. It explores a certain domain of quantum vacua via the gauge theoretic fluxes underlying a strong-weak coupling duality. On the other hand, there are theoretical attempts  \cite{steinacker,carlip,padmanabhan} to investigate an emergent quantum gravity underlying Einstein vacuum. An emergent gravity has also been conjectured to arise due to the statistical behavior of microscopic degrees encoded on a holographic screen \cite{verlinde}.

\sp
\noindent
In the past there have been attempts to construct extremal geometries on a BPS $D$-brane in various dimensions \cite{seiberg-witten}-\cite{kahle-minasian}. They correspond to the near horizon black holes primarily described in a ten dimensional type IIA or type IIB superstring theory. Importantly a global NS two form in the string bulk couples to an electromagnetic field on a $D$-brane to form a non-linear $U(1)$ gauge invariant combination \cite{seiberg-witten}. However a non-constant NS two form remains in the string bulk and hence does not play a role on a BPS D-brane. Nevertheless a closed string world-sheet conformal symmetry in presence of a NS two form of a two form lead to the vanishing of beta function equations and reassures a superstring effective action in ten dimensions. Interestingly a NS two form modifies the covariant derivative and is known to describe a torsion in an effective string theory \cite{candelasHS,freed}.

\sp
\noindent
In the context an effective torsion curvature formalism underlying a two form in a $U(1)$ gauge theory on a $D_4$-brane was obtained by the the authors in the recent past \cite{spsk,kpss3}. The five dimensional de sitter geometries obtained in the formalism were shown to be sourced by a geometric torsion. The formalism was exploited to describe a large number of tunneling vacua including de Sitter and anti de Sitter black holes in five dimensions on a vacuum created pair of $(D{\bar D})$-brane at the Big Bang. The $U(1)$ gauge theory, underlying an effective curvature, allows a perturbative coupling of a global NS two form to a gauge theoretic torsion $H_3$ on a $D_4$-brane. Interestingly some of the five dimensional torsion geometries on $S^1$ were shown to describe torsion free black holes on a vacuum created $D_3$-brane within a pair \cite{spsk-RN}. Two of the brane universes have respectively been identified with a Reissner-Nordstrom (RN) and a Schwarzschild black hole in Einstein gravity. Subsequently the second order formalism was explored to obtain quantum Kerr geometries in five dimensions \cite{spsk-kerr}. The torsion contribution was argued to be negligibly small in a low energy limit. Hence the torsion potential was ignored to obtain a Kerr black hole in a five dimensional brane universe. In the paper we investigate some of the plausible Kerr geometries on a vacuum created $D_3$-brane universe within a pair using the formalism. More over the effective curvature scenario has been applied to various other interesting developments including a quintessence axion in string cosmology by the authors \cite{kpss1}-\cite{pssk}.

\sp
\noindent
In particular a non-trivial space-time was argued to began on a vacuum created on a pair of $(D{\bar D})_3$-brane at an even horizon by a two form in a $U(1)$ gauge theory on a $D_4$-brane. A two form gauge theory is Poincare dual to an one form dynamics on a $D_4$-brane. Thus the pair production idea is inspired by the established phenomenon of Hawking radiation at the event horizon of a background black hole by a photon in a $U(1)$ gauge theory \cite{hawking}. It may imply that a the gauge theory on a $D_4$-brane may alternately be viewed through a vacuum created pair of $(D{\bar D})_3$-brane and an anti $D_3$-brane in a five dimensional effective curvature formalism. An extra fifth dimension, transverse to a vacuum created pair of $(D{\bar D})_3$-brane, has been shown to play a significant role. It is believed take into account the perturbative closed string vacuum between a $D_3$-brane and an anti $D_3$-brane. In other words the formalism underlying a geometric torsion may seen to describe a low energy perturbative closed string vacuum in presence of a non-perturbative $D$-brane and an anti $D$-brane corrections \cite{spsk-kerr,psskk,psskQ,pssk}. In the context a flat metric has been shown to be associated with a quantum correction. It indeed reassures the absence of closed string modes on a $D$-brane. In particular a non-perturbative correction is shown to be sourced by a non-linear $U(1)$ charge which possesses its origin in a geometric torsion. As a result a vacuum created brane universe may be approximated with the Einstein vacuum in presence of a non-perturbative quantum correction. The presence of a fifth dimension hidden to a 4D non-extremal quantum vacua may allow one to imagine the existence of another universe described by an anti-BPS brane. In fact a generic torsion curvature has been argued to began with a vacuum created pair of $(D{\bar D})$-instanton at the Big Bang by a two form \cite{spsk}. Since the formalism evolves with an intrinsic torsion, a $(D{\bar D})$-brane is inevitable to nullify the torsion while establishing a correspondence with a known vacuum in Einstein gravity.

\sp
\noindent
The underlying setup may be illustrated by considering an effective open string metric on an anti $D_4$-brane underlying a global NS two form on its world-volume. The presence of a vacuum created anti $D$-brane sourced by a geometric torsion in the formalism imply the existence of a $D$-brane at a transverse distance. Thus the dynamics on a $D_4$-brane, underlying the scenario of a pair of brane/anti-brane, may be described by a two form in presence of a background (open string) metric in a $U(1)$ gauge theory. The non-linear quanta of a two form, or its Poincare dual one form, is argued to vacuum create a pair of $(D{\bar D})_3$-brane at the event horizon of an effective black hole established on an anti $D_4$-brane. Needless to mention that a BPS brane and an anti BPS brane together breaks the supersymmetry and describes a non-BPS brane configuration. It may be noted that the formalism evolves with two independent two forms. One of them is a dynamical two form which is Poincare dual to an one form on a $D_4$-brane. The other one is a global NS two form on an anti $D_4$-brane which is known to describe an effective open string metric and hence an effective gravity may seen to emerge on an anti $D_4$-brane. In other words a two form dynamics in a $U(1)$ gauge theory on a $D_4$-brane is explored in presence of a background black hole. Three local degrees of a two form on $S^1$ is described by an axionic scalar on an anti $D_3$-brane and a non-linear one form on a $D_3$-brane.

\sp
\noindent
In the paper we obtain a number of quantum vacua on a created pair of $(D{\bar D})_3$-brane primarily sourced by a two form on a $D_4$-brane.
We construct the quantum Kerr-Newman geometries leading to a quantum black hole in four dimensions on a non-BPS brane in a superstring theory. 
The brane geometries in a certain regime are analyzed for their mass, angular velocity and charge if any. They are shown to map to Einstein vacua in string theory. A gauge choice for a two form leading to a vanishing torison in a generalized space-time curvature theory ensures Riemannian geometry underlying a vacuum $T_{\mu\nu}=0$. It is argued that the background fluctuations in $B_2$ on a $D$-brane may have their origin in a dynamical two form described by a ten dimensional effective closed string action. In particular a five dimensional quantum Kerr black hole with two angular momenta have 
been obtained by the authors using a generalized curvature \cite{spsk-kerr}.  A quantum Kerr obtained in the formalism is analyzed in a low energy limit to describe a typical Kerr black hole in a five dimensional Einstein gravity \cite{sakaguchi}. Interestingly a number of rotating black holes in various dimensions are obtained in the folklore of Einstein gravity \cite{myers-perry}-\cite{nikola}.

\sp
\noindent
We observe that an event horizon in a 4D quantum Kerr(Newman) black hole is independent of its charge sourced by an one form. The charge independence identifies a magnetically charged quantum vacuum with an electrically charged one. In fact the quantum Kerr(Newman) black hole is argued to describe a degenerate Kerr vacua. A low energy limit removes the degeneracy in the Kerr and relates to a classical Kerr black hole in Einstein vacuum. Interestingly a (semi) classical Kerr black hole formally retains the causal patches of the quantum Kerr. It may imply that an exact geometry in perturbation theory validates a non-perturbative construction realized through a geometric torsion on a non-BPS brane. We show that a magnetic non-linear charge can be absorbed by a renormalized mass to describe a quantum Schwarzschild black hole for an $S_2$-restoring geometry. In a low energy limit the quantum black hole may be identified with the Schwarzschild black hole in Einstein vacuum. On the other hand an electric non-linear charge underlying the degeneracies in a quantum Kerr, for an $S_2$-restoring geometry, is shown to describe a RN quantum black hole in four dimensions. Its low energy vacuum is shown to describe a RN-black hole in Einstein-Maxwell theory. The renormalization of the mass in a RN quantum black hole has been invoked to argue for a plausible tunneling to a Schwarzschild quantum black hole.

\sp
\noindent
We plan the paper as follows. A non-perturbative geometric torsion leading to an effective curvature underlying a non-linear $U(1)$ theory on a $D_4$-brane is revisited in section 2. We work out the quantum Kerr(Newman) geometries on a vacuum created pair of $(D{\bar D})_3$-brane in section 3. The degenerate Kerr(Newman) vacua are analyzed in a limit to obtain Reissner-Nordstrom and Schwarzschild quantum black holes in section 4. We explore a low energy limit leading to some of the semi-classical Einstein vacua in section 5. We conclude by summarizing the results obtained in an effective five dimensional curvature formalism underlying the quantum effects and outlining some of the future perspectives in section 6.

\section{A non-perturbative setup sourced by a two form}
\subsection{Geometric torsion on a ${\mathbf{{D}_4}}$-brane}
A BPS brane carries an appropriate RR-charge and is established as a non-perturbative dynamical object in a ten dimensional type IIA or IIB superstring theories. In particular, a $D_4$-brane is governed by a supersymmetric gauge theory on its five dimensional world-volume. However, we restrict to the bosonic sector and begin with the $U(1)$ gauge dynamics in presence of a constant background metric $g_{\mu\nu}$ on a $D_4$-brane. A linear one form dynamics is given by
\be
S_{\rm A}= -{1\over{4C_1^2}}\int d^5x\ {\sqrt{-g}}\ F^2\ ,\label{gauge-1}
\ee 
where $C_1^2=(4\pi^2g_s){\sqrt{\alpha'}}$ denotes the gauge coupling. Remarkably a non-linear $U(1)$ gauge symmetry is known to be preserved in an one form theory in presence of a constant two form on a $D$-brane. In the past there were several attempts to approximate a non-linear $U(1)$ gauge dynamics by Dirac-Born-Infeld action coupled to Chern-Simmons on a BPS D-brane \cite{bengt-1,bengt-2}. The BPS brane dynamics is known to describe an extremal black hole which corresponds to a near horizon geometry in a string theory.

\sp
\noindent
In the context a non-linear $U(1)$ gauge dynamics may also be re-expressed in terms of a two form alone on a $D_4$-brane which is Poincare dual to an one form. The duality allows one to address the one form non-linear gauge dynamics in presence of a constant two form on a $D_4$-brane. A Poincare duality interchanges the metric signature between the original and the dual. The two form gauge theory on a $D_4$-brane may be given by
\bea
S_{\rm B}=- {1\over{12C_2^2}}\int d^5x\ {\sqrt{-g}}\ H^2\ ,\label{gauge-3}
\eea
where $C_2^2=(8\pi^3g_s){\alpha'}^{3/2}$ denotes a gauge coupling.
The local degrees in two form on a $D_4$-brane have been exploited to construct an effective space-time curvature scalar ${\tilde{\cal K}}$ in a second order formalism \cite{kpss1}-\cite{spsk-kerr}. The curvature is primarily sourced by a two form gauge theory on a $D_4$-brane.
Generically an irreducible scalar ${\tilde{\cal K}}$ governs a geometric torsion ${\cal H}_{\mu\nu\lambda}$ which is primarily described by a gauge theoretic torsion $H_{\mu\nu\lambda}=3\nabla_{[\mu}B_{\nu\lambda ]}$ on a $D_4$-brane. Unlike the extremal brane geometries, the dynamics of a geometric torsion on an effective curvature formalism addresses some of the non-extremal quantum vacua in string theory. In fact the emergent black holes are described by a pair of brane and anti-brane separated by a transverse dimension. The $(D{\bar D})_4$-pair breaks the supersymmetry and may describe a non-BPS brane in string theory.

\sp
\noindent
A priori the required modification to incorporate a geometric notion may be viewed via a modified covariant derivative defined with a completely antisymmetric gauge connection: ${H_{\mu\nu}}^{\lambda}$. The appropriate derivative may be given by
\be
{\cal D}_{\lambda}B_{\mu\nu}=\nabla_{\lambda}B_{\mu\nu} + {1\over2}{{H}_{\lambda\mu}}^{\rho}B_{\rho\nu} -{1\over2} {{H}_{\lambda\nu}}^{\rho}B_{\rho\mu}\ .\label{gauge-5}
\ee
Under an iteration $H_3\rightarrow {\cal H}_3$ the geometric torsion in a second order formalism may be defined with all order corrections in $B_2$ in a gauge theory. Formally a geometric torsion may be expressed in terms of gauge theoretic torsion and its coupling to two form. It is given by
\bea
{\cal H}_{\mu\nu\lambda}&=&3{\cal D}_{[\mu}B_{\nu\lambda ]}\nonumber\\
&=&3\nabla_{[\mu}B_{\nu\lambda ]} + 3{{\cal H}_{[\mu\nu}}^{\alpha}
{B^{\beta}}_{\lambda ]}\ g_{\alpha\beta}
\nonumber\\ 
&=&H_{\mu\nu\lambda} + \left ( H_{\mu\nu\alpha}{B^{\alpha}}_{\lambda} + \rm{cyclic\; in\;} \mu,\nu,\lambda \right )\ +\ H_{\mu\nu\beta} {B^{\beta}}_{\alpha} {B^{\alpha}}_{\lambda} + \dots \;\ .\label{gauge-6}
\eea 
An exact covariant derivative in a perturbative gauge theory may seen to define a non-perturbative covariant derivative in a second order formalism. Thus 
a geometric torsion constructed with a modified covariant derivative (\ref{gauge-5}) may equivalently be described by an appropriate curvature tensor ${\tilde{\cal K}}_{\mu\nu\lambda\rho}$ which has been worked out by the authors \cite{spsk}. Explicitly the effective curvature tensors are given by
\bea
{{\tilde{\cal K}}_{\mu\nu\lambda}{}}^{\rho}&=&{1\over2}\partial_{\mu}{{\cal H}_{\nu\lambda}}^{\rho} -{1\over2}\partial_{\nu} {{\cal H}_{\mu\lambda}}^{\rho} 
+ {1\over4}{{\cal H}_{\mu\lambda}}^{\sigma}{{\cal H}_{\nu\sigma}}^{\rho}-{1\over4}{{\cal H}_{\nu\lambda}}^{\sigma}{{\cal H}_{\mu\sigma}}^{\rho}\ ,\nonumber\\ 
{\tilde{\cal K}}_{\mu\nu}&=& -\left (2\partial_{\lambda}{{\cal H}^{\lambda}}_{\mu\nu} +
{{\cal H}_{\mu\rho}}^{\lambda}{{\cal H}_{\lambda\nu}}^{\rho}\right )
\nonumber\\
{\rm and}\qquad {\tilde{\cal K}}&=& -{1\over{4}}{\cal H}_{\mu\nu\lambda}{\cal H}^{\mu\nu\lambda}
\ .\label{gauge-7}
\eea
The fourth order tensor is antisymmetric within a pair of indices, $i.e.\ \mu\leftrightarrow\nu$ and $\lambda\leftrightarrow\rho$, which retains a property of Riemann tensor ${R_{\mu\nu\lambda}}^{\rho}$. However the effective curvature ${{\tilde{\cal K}}_{\mu\nu\lambda}{}}^{\rho}$ do not satisfy the symmetric property, under an interchange of a pair of indices, as in Riemann tensor. Nevertheless, for a constant torsion the 
generic tensor: ${{\tilde{\cal K}}_{\mu\nu\lambda}{}}^{\rho}\rightarrow {R_{\mu\nu\lambda}}^{\rho}$. As a result, the effective curvature constructed in a non-perturbative formalism may be viewed as a generalized curvature tensor. It describes the propagation of a geometric torsion in a second order formalism.
\subsection{Emergent metric fluctuations }
A geometric torsion ${\cal H}_3$ in a second order formalism may seen to break the $U(1)$ gauge invariance of a two form in the underlying gauge theory.  Nevertheless an emergent notion of metric fluctuation, sourced by a two form local degrees, restores gauge invariance in a generalized irreducible space-time curvature ${\tilde{\cal K}}$. The non-perturbative fluctuations, underlying a $U(1)$ gauge invariance, turn out to be governed by the fluxes and are given by
\be
f_{\mu\nu}^{nz}= C\ {\cal H}_{\mu\alpha\beta}\ {{\bar{\cal H}}^{\alpha\beta}{}}_{\nu}\ 
\approx C\ H_{\mu\alpha\beta}\ {{{\bar{\cal H}}}^{\alpha\beta}{}}_{\nu}\ ,\label{gauge-7}
\ee
where $C$ is an arbitrary constant and ${\bar{\cal H}}_{\mu\nu\lambda}= (2\pi\alpha'){\cal H}_{\mu\nu\lambda}$. The generalized curvature tensor may also be viewed though a geometric field strength ${\cal F}_2$ which is Poincare dual to ${\cal H}_3$ on a $D_4$-brane. Then, a geometric ${\cal F}_2$ may be given by
\bea
{\cal F}_{\alpha\beta}&=&{\cal D}_{\alpha}A_{\beta} -{\cal D}_{\beta}A_{\alpha}\nonumber\\
&=&\left ( {\cal F}^z_{\alpha\beta} + {{\cal H}_{\alpha\beta}}^{\delta}A_{\delta}\right )\ ,\label{gauge-8}
\eea
where ${\bar{\cal F}}^z_{\alpha\beta}= (2\pi\alpha'){\cal F}^z_{\alpha\beta}=\left ({\bar F}_{\alpha\beta}+ B^z_{\alpha\beta}\right )$ is defined with a global NS $B^z_{\mu\nu}$ on a $D_4$-brane. It signifies a non-linear electromagnetic field and is gauge invariant under a non-linear $U(1)$ transformations \cite{seiberg-witten}. Apparently a non-zero $H_3$ seems to break the $U(1)$ gauge invariance. Nevertheless, an action defined with a lorentz scalar ${\cal F}^2$ may seen to retain the gauge invariance with the help of an emerging notion of metric fluctuations in the formalism. Then, the fluctuations (\ref{gauge-7}) in its dual description may be given by
\be
f_{\mu\nu}^{nz}={\tilde C} {\bar{\cal F}}_{\mu\alpha}{{\bar{\cal F}}^{\alpha}{}}_{\nu}\ ,\label{gauge-9}
\ee
where ${\tilde C}$ is an arbitrary constant. The dynamical fluctuations in eqs.(\ref{gauge-7}) and (\ref{gauge-9}) modify the constant metric
on a $D_4$-brane. Then, the emergent metric on a $D_4$-brane becomes 
\bea
G_{\mu\nu}&=&\left ( G^z_{\mu\nu}\ +\ C\ {\bar{\cal H}}_{\mu\lambda\rho}\ {{\cal H}^{\lambda\rho}{}}_{\nu}\right )\nonumber\\
&=&\left ( G^z_{\mu\nu}\ +\ {\tilde C}\ {\bar{\cal F}}_{\mu\lambda}{{\bar{\cal F}}^{\lambda}{}}_{\nu}\right )
\ .\label{gauge-10}
\eea 
The fluxes in a bilinear combination are gauge invariant. The emergent metric sourced by the fluxes a priori seems to be unique. 
However an analysis reveals that the emergent metric may not be unique due to the coupling of $B_2$-potential to $H_3$ underlying a geometric torsion ${\cal H}_3$. In other words $B_2$-fluctuations do play a significant role to define the emergent geometries on a vacuum created brane universe. They lead to a generalized notion of metric on a $D$-brane. It may be given by
\be
G_{\mu\nu}=\left ( g_{\mu\nu}\ -  B_{\mu\lambda}{B^{\lambda}}_{\nu}\ +\ C\ {\bar{\cal H}}_{\mu\lambda\rho}\ {{\cal H}^{\lambda\rho}{}}_{\nu}
+\ {\tilde C}\ {\bar{\cal F}}_{\mu\lambda}{{\bar{\cal F}}^{\lambda}{}}_{\nu}\right )
\ .\label{gauge-11}
\ee
The background fluctuations arising out of the non-dynamical components in $B_{\mu\nu}$ may seen to deform the brane geometries significantly. They may lead to a large number of vacua and may correspond to the landscape quantum geometries in the formalism. The background fluctuations in two form 
may have their origin in a higher dimensional gauge theoretic torsion $H_3$. They may couple to an electro-magnetic field in higher dimensions to define a gauge invariant non-linear ${\cal F}^z_2$.

\subsection{Non-perturbative space-time curvature}
The gauge dynamics on a $D_4$-brane in presence of gauge connections, may be approximated by an irreducible generalized curvature theory in a second order formalism. A priori the effective curvature may seen to describe a geometric torsion dynamics on an effective $D_4$-brane \cite{spsk}. A geometric construction of a torsion in a non-perturbative formalism is inspiring and may provoke thought to unfold certain aspects of quantum gravity. Generically the action may be given by
\be
S_{\rm D_4}^{\rm eff}= {1\over{3C_4^2}}\int d^5x {\sqrt{-G}}\ \left ( {\tilde{\cal K}}- \Lambda\right )\ ,\label{gauge-12}
\ee
where $C_4^2=(8\pi^3g_s){\alpha'}^{3/2}$ is a constant and $G=\det G_{\mu\nu}$. The cosmological constant ${\Lambda}$, in the geometric action is sourced by a global NS two form in the theory. With $\kappa^2=(2\pi)^{5/2}g_s\alpha'$ a generalized curvature dynamics on $S^1$, underlying an effective  $D_3$-brane \cite{spsk-RN}, may appropriately be given by
\be
S_{\rm D_3}^{\rm eff}= {1\over{3\kappa^2}}\int d^4x {\sqrt{-G}}\ \left ( {\cal K}\ - {\Lambda}\ -\ {3\over4} {\bar{\cal F}}_{\mu\nu}
{\cal F}^{\mu\nu} \right )\ .\label{gauge-13}
\ee
The curvature scalar ${\cal K}$ is sourced by a dynamical two form in a non-linear $U(1)$ gauge theory. The field strength ${H_{\mu\nu}}^{\lambda}$ is governed by the equations of motion of a two form in presence of a flat background metric $g_{\mu\nu}$ in a $U(1)$ gauge theory. Explicitly the $B_2$ field equations of motion are given by
\be
\partial_{\lambda}H^{\lambda\mu\nu} + {1\over2} g^{\alpha\beta}\partial_{\lambda}\ g_{\alpha\beta}\ 
H^{\lambda\mu\nu}=0\ . \label{gauge-KerrD-1}
\ee
The field strength $H_3\rightarrow {\cal H}_3$ is appropriately modified to describe a propagating torsion in four dimensions underlying a second order formalism. The fact that a torsion is dual to an axion on an effective $D_3$-brane ensures one degree of freedom. In addition ${\cal F}_{\mu\nu}$ describes a geometric one form field with two local degrees on an effective $D_3$-brane. A precise match among the (three) local degrees of torsion in ${\cal K}^{(5)}$ on $S^1$ with that in ${\cal K}$ and ${\cal F}_{\mu\nu}$ reassure the absence of a dynamical dilaton field in the frame-work. The result is consistent with the fact that a two form on $S^1$ does not generate a dilaton field. The equation of motion for the one form is defined with an appropriate covariant derivative. It is given by
\be
{\cal D}_{\lambda}{\cal F}^{\lambda\nu}=0\ ,\qquad {}\label{gauge-KerrD-2}
\ee
\noindent
where ${\cal F}_{\mu\nu}=\left ({\cal D}_{\mu}A_{\nu}-{\cal D}_{\nu}A_{\mu}\right )$ and ${\cal D}_{\mu}A_{\nu}=\left (\nabla_{\mu}A_{\nu} + {1\over2}{{\cal H}_{\mu\nu}}^{\lambda}A_{\lambda}\right )$. The energy-momentum tensor $T_{\mu\nu}$ is computed in a gauge choice: 
\be
{\bar{\cal F}}_{\mu\nu}{\cal F}^{\mu\nu}= {4\over{\pi\alpha'}} + {4\over3}\left ({\tilde{\cal K}}-\Lambda\right )\ .\label{gauge-KerrD-3}
\ee
Interestingly the $T_{\mu\nu}$ in a non-linear gauge theory on a $D_3$-brane incorporates an emergent metric in a generalized curvature theory, $i.e.\ (2\pi\alpha')T_{\mu\nu}=G_{\mu\nu}$. The gauge choice ensures that the $T_{\mu\nu}$ sources a nontrivial emergent geometry underlying a non-linear $U(1)$ gauge theory. The covariant derivative satisfies ${\cal D}_{\lambda}{G}_{\mu\nu}=0$. Thus, an emergent metric in the framework uniquely fixes the covariant derivative. It implies ${\cal D}_{\mu}T^{\mu\nu}=0$, which in turn incorporates a conserved charge $Q$ in the formalism. The effective curvature theory may formally be viewed as a non-linear $U(1)$ gauge theory. The energy-momentum tensor may be given by
\bea
T_{\mu\nu}&=&{1\over{6}}\left ({\Lambda}-{\tilde{\cal K}}\right ){G}_{\mu\nu} - {1\over{8C\pi\alpha'}}f_{\mu\nu}^{nz}\nonumber\\
&=&{1\over6}\left ( {\Lambda}-{\tilde{\cal K}}\right )G^z_{\mu\nu} + \left (\Lambda-{{\tilde{\cal K}}\over{6}} - {1\over{8C\pi\alpha'}}\right )f_{\mu\nu}^{nz}\ . \label{gauge-13}
\eea
The trace of energy-momentum tensor on a $D_3$-brane becomes
\be
T={1\over3}\left ({\tilde{\cal K}}+{2\Lambda}\right )\ .\label{gauge-14}
\ee
It ensures that a vacuum, $i.e.\ T=0$, may be defined in a gauge choice: (${\tilde{\cal K}}=0$ and $\Lambda=0$) or (${\tilde{\cal K}}=-2\Lambda$).
We consider a gauge choice:
\be
{\Lambda}=\left ( {3\over{\pi\alpha'}}\right )\ +\ {\tilde{\cal K}}\ .\label{gauge-141}
\ee
Then a $T_{\mu\nu}$ in the gauge choice sources a generic ($C\neq 1/4$ and ${\tilde C}\neq -1/2$) emergent metric on an arbitrary dimensional $D$-brane underlying an effective curvature ${\tilde{\cal K}}^{(5)}$ . It is given by
\bea
T_{\mu\nu}&=&\left ( {{G^z_{\mu\nu}}\over{2\pi\alpha'}}\ +\ \left [ C-{1\over4}\right ]\ {\cal H}_{\mu\lambda\rho} {{\cal H}^{\lambda\rho}}_{\nu}
+\ \left [ {\tilde C}+{1\over2}\right ]\ {\bar {\cal F}}_{\mu\lambda} {{\cal F}^{\lambda}}_{\nu}\right )\nonumber\\
&=&\left ( {{G_{\mu\nu}}\over{2\pi\alpha'}}\ -\ {1\over4} {\cal H}_{\mu\lambda\rho} {{\cal H}^{\lambda\rho}}_{\nu}
\ +\ {1\over2} {\bar{\cal F}}_{\mu\lambda} {{\cal F}^{\lambda}}_{\nu}\right )\ .\label{gauge-31}
\eea
Thus the $T_{\mu\nu}$ in a gauge theory sources the dynamics of a torsion in a generalized curvature theory. A higher dimensional $T_{\mu\nu}$ can source a lower dimensional background fluctuations in two form on a brane. A gauge choice (\ref{gauge-141}) in eq.(\ref{gauge-KerrD-3}) ensures ${\cal F}^2=0$ and hence the metric fluctuation $\left ({\cal H}_{\mu\lambda\rho} {{\cal H}^{\lambda\rho}}_{\nu}\right )$ becomes significant on a $D_4$-brane.
Thus the non-trivial flux in a gauge choice leads to a dynamical geometric torsion underlying a generalized curvature ${\tilde{\cal K}}$ on a $D_4$-brane.

\section{${\mathbf{S_2}}$-deformed quantum vacua on a ${\mathbf{(D{\bar D})_3}}$-brane}
In this section, we obtain the quantum Kerr geometries constructed on a non BPS brane in four dimensions in presence of a non-linear charge $Q$.  Interestingly a quantum Kerr-Newman black hole, on a non-BPS brane in a superstring theory, may seen to describe a number of vacua in various lower energy scales to Planckian energy. Arguably, the vacuum geometries tunnel among themselves at an intermediate energy scale. The brane geometries obtained in a non-perturbative formalism are analyzed in a low energy limit to realize, a Kerr-Newman, a Kerr, a Reissner-Nordstrom and a Schwarzschild, black holes in Einstein vacuum. In fact, we address some of the rotating quantum black holes obtained on an effective $D_3$-brane underlying a generalized curvature ${\tilde{\cal K}}$ on $S^1$. 

\subsection{Flat metric}
We use Boyer-Lindquist coordinates on a $D_3$-brane underlying a Minkowski space-time. The range: $0<$$\phi$$\le$$2\pi$ and $0$$<$$\theta$$\le$$\pi$ completely specify the flat space for the line-element. Now, we set $(2\pi\alpha')=1$ in the paper. The cartesian coordinates may be defined by a spheroidal coordinate system. They are:
\be
x={\sqrt{r^2+a^2}}\left (\sin\t \cos\phi\right ),\; y={\sqrt{r^2+a^2}}\left (\sin\t \sin\phi\right )\ ,\ z=r\ \cos\t\ ,\label{KNewman-brane-31}
\ee
where $a$ is an arbitrary constant and shall be identified with a $S_2$ symmetry breaking background parameter. The coodinates ensure a circle in $xy$-plane with a varying $z$-coordinate for $\t\neq (0,\pi )$. It is given by: $(x^2+y^2) = (r^2+a^2)\sin^2\t$ and $z^2 = r^2\cos^2\t$. Nevertheless, for $\phi=\pi, 2\pi$ and $\t\neq (0,\pi )$, the coordinates ensure a circle in $xz$-plane and is given by: $(x^2+z^2) = r^2 + a^2\sin^2\t$ and $y=0$. Similarly for 
$\phi=\pi/2, 3\pi/2$ and $\t\neq (0,\pi )$, a circle is defined in $yz$-plane: $(y^2+z^2) = r^2 + a^2\sin^2\t$ and $x=0$. Generically, all three circle equations satisfy: 
\be
x^2 +y^2 + z^2= r^2 + a^2\sin^2\t\ .\label{KNewman-brane-312}
\ee
The equations in $xy$-, $xz$- and $yz$-planes on the equator, respectively, define a ring of radius ${\sqrt{r^2+a^2}}$ for $z=0$, $y=0$ and $x=0$. The Minkowski vacuum on a $D_3$-brane defined with a flat metric $g_{\mu\nu}$ in Boyer-Lindquist coordinates is given by
\bea
ds^2_{\rm flat}&=&-\ dt^2\ +\ {{\rho_a^2}\over{\tri_a}}\ dr^2\ +\ \rho_a^2\ d\t^2\ +\ \tri_a\sin^2\t \ d\phi^2\nonumber\\
&=&-\ dt^2\ +\ {{\rho_a^2}\over{\tri_a}}\left (dr^2\ +\ \tri_a\ d\Omega^2\right )\ +\ a^2\sin^4\t \ d\phi^2\nonumber\\
&=&-\ dt^2 + dr^2 + r^2\ d\Omega^2\nonumber\\
&&\qquad\;\; +\ a^2\left (- {{\sin^2\t}\over{\tri_a}} dr^2 + \cos^2\t\ d\t^2 + \sin^2\t \ d\phi^2\right )
\ ,\label{KNewman-brane4}
\eea
$${\rm where}\quad \tri_a=(r^2+a^2)\quad {\rm and}\;\ \rho^2_a=(r^2+a^2\cos^2\t )\ .$$ 
In a limit $\rho_a^2\rightarrow \tri_a$, the first expression assures a flat $S_2$-symmetric vacuum on a $D_3$-brane. It may seen to simplify the quantum geometry without changing its characteristic properties. Generically the limit may as well describes the geometry at poles for $a\neq 0$. The second and third expressions confirm that the $S_2$-symmetry in a vacuum is broken by a perturbation parameter `$a$'. 
Contrary to a forbidden limit for the effective radius $\rho\rightarrow 0$ on a $D_4$-brane \cite{spsk-kerr}, the limit $\rho_a\rightarrow 0$ turns out to be an allowed on the equator for $r\rightarrow 0$ on a $D_3$-brane. In the limit, the metric possesses a coordinate ring singularity on a equatorial plane.

\subsection{Gauge field ansatz}
A two form on a $D_3$-brane satisfies the equation of motion (\ref{gauge-KerrD-1}) in a non-linear $U(1)$ gauge theory. A two form ansatz leading to 
a family of Kerr vacua in Einstein gravity is worked out. In presence of a spherical symmetry breaking perturbation parameter $a$, the two form 
ansatz may be expressed as:
\bea
B_{tr}&=&\left ( {{2M}\over{\tri_a}}\right )^{1/2}\qquad\qquad\qquad\qquad\qquad\qquad\qquad\qquad\; {}\nonumber\\
&=&B_{tr}^{(a=0)}\left ( 1 + {{a^2}\over{r^2}}\right )^{-1/2}\nonumber\\
{\rm and}\qquad B_{r\t}&=&\rho_a\left ({{a^2\sin^2\t + 2M}\over{\tri_a}} - {{a^2\sin^2\t -2M}\over{\rho_a^2}}\right )^{1/2}\nonumber\\
&=&B_{r\t}^{(a=0)} \left ( 1 - {{a^2\sin^2\t}\over{2(r^2+a^2)}}\left [ 1 + {{a^2\sin^2\t}\over{2M}}\right ]\right )^{1/2}\ ,\label{KNewman-brane1}
\eea
where $M$ is an arbitrary constant and shall be identified with a mass with a lower cutoff $M\ge a^2$ in quantum gravity. 

\sp
\noindent
On the other hand, the one form ansatz may be given by
\bea
A_t&=&-\ {{Qr}\over{\rho_a^2}}\qquad\qquad\qquad\qquad\qquad\qquad {}\nonumber\\
&=&A_t^{(a=0)} \left ( 1 + {{a^2\cos^2\t}\over{r^2}}\right )^{-1}\nonumber\\
{\rm and}\quad A_{\phi}&=&{{aQr \sin^2\t}\over{\rho_a^2}}\nonumber\\
&=&-A_t^{(a=0)} \left ({{ar^2 \sin^2\t}\over{r^2+a^2\cos^2\t}}\right )\ ,\label{KNewman-brane2} 
\eea
where an arbitrary constant $Q$ signifies a non-linear electric charge presumably with a lower cutoff $Q$$\ge$$a^2$. Generically a lower bound in $M$ and $Q$ are enforced by their non-linearity, which can not be gauged away completely in a gauge theory. It shows that the $A_{\phi}$ component is sourced by the $A_t$ in presence of the background parameter $a$. Thus, a magnetic field is generated via the spherical symmetry breaking parameter `$a$' from an electric field. We shall see that the intermingle, of an electric with a magnetic field, phenomenon may well be described with a subtlety. Presumably, the parameter incorporates an euclidean notion of time within an ergo sphere of an emergent black hole. The non-zero components of field strength are worked out to yield: 
\bea
{\cal F}_{tr}&=&-{Q\over{\rho_a^4}} \left (r^2-a^2\cos^2\t\right )\ ,\nonumber\\
{\cal F}_{t\t}&=&{{a^2Qr}\over{\rho_a^4}} \sin2\t\ ,\nonumber\\
{\cal F}_{r\phi}&=&-{{aQ}\over{\rho_a^4}}\left (r^2-a^2\cos^2\t\right )\sin^2\t\nonumber\\
{\rm and}\quad {\cal F}_{\t\phi}&=&{{aQr\tri_a}\over{\rho_a^4}}\sin2\t\ .\label{KNewman-brane3}
\eea
In a limit $a\rightarrow 0$ the electromagnetic field in the geometric framework may only be described by a non-linear electric field. In the gauge choice the gauge theoretic torsion $H_3=0$ and hence the geometric torsion ${\cal H}_3=0$. The gauge ansatz freezes the local degrees of torsion on a $D_3$-brane within a vacuum created pair of brane/anti-brane. In the case the brane dynamics is solely contributed by an one form in four dimensions. A nontrivial geometric field strength reduces in the gauge choice, $i.e.\ {\cal F}_2\rightarrow {\cal F}_2^z$, and is defined with a non-linear charge $Q$. The non-linearity in charge $Q$ is due to a global NS two form \cite{seiberg-witten}. A non-vanishing ${\cal F}^z_2$ implies a non-trivial energy-momentum tensor in the $U(1)$ gauge theory on a $D_3$-brane. However a vanishing trace of energy momentum tensor in 4D may hint for a vacuum solution in Einstein gravity. Thus a conserved non-linear $U(1)$ charge $Q$ is defined in absence of a torsion on a $D_3$-brane. The gauge field equations of motion in the case leads to $\nabla_{\mu}J^{\mu}=0$ for an appropriate conserved current $J^{\mu}$ and defines a conserved charge in the formalism.
 
\subsection{Causal geometric patches}
The $B_2$-fluctuations in presence of ${\cal F}^z_2$ may seen to source a non-trivial emergent metric on an effective $D_3$-brane. For ${\tilde C}=\pm 1$ the emergent metric in the gauge choice on a $D_3$-brane reduces to yield: 
\be
G_{\mu\nu}\rightarrow \left ( g_{\mu\nu}\ -\ B_{\mu\alpha}g^{\alpha\beta}B_{\beta\nu}\ \pm\ {\bar{\cal F}}^z_{\mu\alpha}g^{\alpha\beta}
{\bar{\cal F}}^z_{\beta\nu}\right )\ .\label{KNewmann-brane5}
\ee
Two geometries for an emergent metric tensor is a choice keeping the generality in mind. Primarily they dictate the quantum geometric corrections and incorporate local degrees to a background metric on a $D_3$-brane. The non-trivial metric components sourced by the gauge fields on a $D_3$-brane are worked out to yield:
\bea
G_{tt}&=&-\left(1-\frac{2M}{\rho_a^2} \pm  {{Q^2}\over{\rho_a^6}}\left [\tri_a-\frac{4r^2a^2\cos^2\t}{\rho_a^2}\right ]\right )\ , \nonumber\\
G_{rr}&=&\left(1+\frac{2M - a^2\sin^2\t}{\rho_a^2} \pm \frac{Q^2}{\rho_a^2\tri_a}\left [ 1- {{4r^2a^2\cos^2\t}\over{\rho_a^4}}\right ]\right)\ ,\nonumber\\ 
G_{\t\t}&=&\rho_a^2 + 2M\left [ 1+{{\tri_a}\over{\rho_a^2}}\right ] + a^2\sin^2\t\left [1- {{\tri_a}\over{\rho_a^2}}\right ]\mp \frac{4a^2Q^2r^2\cos^2\t}{\rho_a^6}\ ,\nonumber\\
G_{\phi\phi}&=&\left ( 1 \mp {{a^2Q^2}\over{\rho_a^6}}\left [\sin^2\t + \frac{4a^2r^2\cos^2\t}{\rho_a^2}\right ]\right )\tri_a\sin^2\t\ ,\nonumber\\ 
G_{t\t}&=&-\frac{\sqrt{2M\tri_a}}{\rho_a}\left (\frac{a^2\sin^2\t + 2M}{\tri_a} - \frac{a^2\sin^2\t -2M}{\rho_a^2}\right )^{1/2}\nonumber\\ 
{\rm and}\quad G_{t\phi}&=&\pm \ {{aQ^2}\over{\rho_a^6}}\tri_a\sin^2\t\ .\label{KNewman-brane6}
\eea
The emergent metric components on a ${\bar D}_3$-brane may be obtained under $r\rightarrow -r$. For simplicity we analyze the geometries in a special limit $\rho_a^2\rightarrow \tri_a$ on a brane. It reduces to yield:
\bea
ds^2&=&-\ \left (1-\frac{2M}{\tri_a} \pm  {{Q^2}\over{\tri_a^2}} \mp \frac{4a^2Q^2r^2\cos^2\t}{\tri_a^4}\right )dt^2\nonumber\\ 
&&+\ \left ( 1 + {{2M-a^2\sin^2\t}\over{\tri_a}} \pm \frac{Q^2}{\tri_a^2} \mp \frac{4a^2Q^2r^2\cos^2\t}{\tri_a^4}\right )\ dr^2\nonumber\\
&&+\ \left (1+\frac{4M}{\tri_a} \mp \frac{4a^2Q^2r^2\cos^2\t}{\tri_a^4}\right )\tri_a\ d\t^2\nonumber\\
&&-\ {{4M{\sqrt{2}}}\over{\sqrt{\tri_a}}} \ dtd\t \pm\ \frac{2aQ^2\sin^2\t}{\tri_a^2}\ dtd\phi \nonumber\\
&&+\ \left ( 1 \mp {{a^2Q^2\sin^2\t}\over{\tri_a^3}} \mp  \frac{4a^4Q^2r^2\cos^2\t}{\tri_a^4}\right )\tri_a\sin^2\t\ d\phi^2\ \ . \label{KNewman-brane65}
\eea
The emergent quantum geometries for $r^2$$>$$2M$ and $r^4$$>$$>$$M^2$ on a brane may be re-expressed by a Kerr black hole in presence of
geometric corrections coupled to a charge $Q$ in a non-linear perturbation theory. They may be given by
\bea
ds^2&=&ds^2_{\rm Kerr}\ \pm\ \frac{Q^2}{\tri_a^2} \left ( -dt^2_{\phi}  + dr^2 \right )\nonumber\\
&&\qquad\quad \mp\ \frac{4a^2Q^2r^2\cos^2\t}{\tri_a^4}\left ( -dt^2 + dr^2 + \tri_a\ d\Omega^2_a\right )\ ,\label{KNewman-brane66}
\eea
where $dt_{\phi}=\left (-dt + a\sin^2\t\ d\phi\right )$, $d\Omega^2_a=\left ( d\t^2 + a^2\sin^2\t\  d\phi^2\right )$ signifies an $S_2$-deformed in presence of the background parameter `$a$'. Interestingly, the quantum (geometric) corrections to $ds^2_{\rm Kerr}$ possess their origin in a flat line-element (\ref{KNewman-brane4}) in a limit $\rho_a\rightarrow \tri_a$. Then the causal patches in eq(\ref{KNewman-brane66}) associated with an electric charge $Q$ are obtained in a limit ${\rho_a^2\rightarrow \tri_a}$. It is given by 
\bea
ds_Q^2&=&\pm \frac{Q^2}{\rho_a^4}\Big [\Big (- dt^2_{\phi}  + {{\rho_a^2}\over{\tri_a}}\ dr^2 \Big )\nonumber\\
&&\ -\ {{4a^2\cos^2\t}\over{\rho_a^2}}\Big (1-{{a^2\cos^2\t}\over{\rho_a^2}}\Big )\Big ( -dt^2 + {{\rho_a^2}\over{\tri_a}}dr^2 + \rho_a^2 d\Omega^2_a\Big )\Big ] .\label{KNewman-brane662}
\eea
It reconfirms a zero curvature in the geometric patches associated with a quantum gravity correction on a pair of $(D{\bar D})_3$-brane in 
a type IIB superstring theory. We shall see that the quantum correction is indeed non-perturbative. A flat metric underlying a quantum correction to an emergent Kerr geometry is remarkable. It may turn out to be a potential tool to explore some of the unresolved issues in quantum gravity. On the other hand, the Kerr line-element in eq(\ref{KNewman-brane66}) may explicitly be given by
\bea
ds^2_{\rm Kerr}&=&-\ \left (1-\frac{2M}{\tri_a} \right )dt^2\ +\ \left ( 1 - {{2M-a^2\sin^2\t}\over{\tri_a}}\right )^{-1} dr^2
\nonumber\\
&&-\ {{4M{\sqrt{2}}}\over{\sqrt{\tri_a}}} \ dtd\t + \left (1+\frac{4M}{\tri_a}\right ) \tri_a\ d\t^2 +\tri_a\sin^2\t\ d\phi^2\ .\label{KNewman-brane67}
\eea
The causal patches characterize a 4D quantum Kerr black hole on a pair of $(D{\bar D})_3$-brane in presence of a fifth dimension. In fact the scale of  an extra dimension distinguishes a low energy vacuum from its quantum vacuum. The Kerr vacuum is sourced only by a two form in a non-linear $U(1)$ gauge theory. Under an interchange of angular coordinates $d\t\leftrightarrow d\phi$ with their appropriate normalizations, the angular velocity $\Omega^{\phi}$ becomes significant at the expense of $\Omega^{\t}$ and leads to a precise Kerr vacuum in Einstein gravity. It reconfirms a Schwarzschild vacuum for $a=0$ in a global scenario underlying a pair of $(D{\bar D})_3$-brane. Presumably an extra dimension transverse to $D_3$-brane and an anti $D_3$-brane in a global scenario does not allow an annihilation of a pair of branes to a BPS brane. The irreversibility of pair creation process may also be argued from the non-linearity in a two form quanta in a superstring theory. 

\sp
\noindent
In fact an emergent Kerr geometry (\ref{KNewman-brane67}) on a $D_3$-brane when comes in contact with that on a ${\bar D}_3$-brane, does not reduce to a flat line-element (\ref{KNewman-brane4}) on a BPS $D_3$-brane. However for $a=0$ an angular velocity $\Omega^{\t}$ on a $D_3$-brane nullifies that on a ${\bar D}_3$-brane and hence reduce to a quantum Schwarzschild black hole. Then the relation in eq(\ref{KNewman-brane66}) takes a form:
\be
ds^2=ds^2_{\rm Sch}\ \pm\ \frac{Q^2}{r^4} \left ( -\ dt^2 \ + \ dr^2\right )\ ,\label{KNewman-brane68}
\ee
where a quantum Schwarzschild $ds^2_{\rm Sch}$ may be obtained from $ds^2_{\rm Kerr}$ in eq(\ref{KNewman-brane67}) for $a=0$. The quantum 4D Schwarzschild black hole obtained on a $(D{\bar D})_3$-brane hints at a small fifth dimension along with a deformed $S_2$-geometry when compared with a
a Schwarzschild black hole in Einstein vacuum. The extra dimension and the deformations are intrinsic to a geometric torsion on a $D_4$-brane in the formalism. In other words the presence of a fifth dimension signifies a propagating torsion in quantum gravity. 

\sp
\noindent
Similarly the quantum geometric patches (\ref{KNewman-brane65}) in a limit $\rho_a^2\rightarrow \tri_a$ may be rearranged in terms of their coupling to $M$ and a charge $Q$ to yield:
\bea
ds^2&=&\left ( -dt^2 + dr^2 + \tri_a\ d\Omega^2\right )\ +\ {{2M}\over{\tri_a}} \left ( dt^2_{\t} + dr^2\right )\ \pm \ {{Q^2}\over{\tri_a^2}} \left ( -dt^2_{\phi} + dr^2\right )\nonumber\\ 
&&\mp\ \ {{4a^2Q^2r^2\cos^2\t}\over{\tri_a^4}} \left (-dt^2 + dr^2 + \tri_a\ d\Omega_a^2
\right )\nonumber\\
&=& \left ( ds^2_{\rm flat}\ +{{2M}\over{\rho_a^2}}\ ds^2_1\right )_{\rho_a^2\rightarrow \tri_a}\nonumber\\
&&\pm\ \left ({{Q^2}\over{\rho_a^4}}\left [ ds_2^2- {{4a^2\cos^2\t}\over{\rho_a^2}}\left ( 1 - {{a^2\cos^2\t}\over{\rho_a^2}}\right )ds^2_3\ \right ]\right )_{\rho_a^2\rightarrow \tri_a}\label{KNewman-brane69}
\eea
where $dt_{\t}=\left (-dt+ \rho_a{\sqrt{2}}\ d\t\right )$ is defined in the limit. The $ds^2_{\rm flat}$ is defined in eq(\ref{KNewman-brane4}) and all the remaining line-elements describe flat causal patches. Explicitly $ds_3^2$ may be given by
$$ds_3^2=\left (-dt^2\ +\ {{\rho_a^2}\over{\tri_a}}\ dr^2\ +\ \rho_a^2\ d\t^2\ +\ a^2\tri_a\sin^2\t\ d\phi^2\right )\ .$$  
A priori the fluxes Weyl scale the flat metric patches on a brane and may not lead to a black hole vacuum in absence of a $B_2$-fluctuation. 
Thus a two form plays a significant role to describe an emergent graviton underlying a non-perturbative geometric torsion in the formalism. On the other hand an one form has been shown to incorporate a non-perturbative quantum correction underlying a flat geometry. Presumably it would describe 
a ``graviton'' within a gauge choice for a vanishing torsion in the formalism.

\sp
\noindent
Furthermore eqs(\ref{KNewman-brane662}) and (\ref{KNewman-brane69}) confirm that an $S_2$-deformation parameter `$a$' incorporates geometric perturbations within a non-perturbative quantum correction. For $a=0$ the quantum geometries (\ref{KNewman-brane69}) reduce drastically on a pair of $(D{\bar D})_3$-brane in a pre-defined regime $r^2$$>$$2M$ with $r^4$$>$$>$$M^2$ and is obtained in eq.(\ref{KNewman-brane68}). Explicitly it is given by
\bea
ds^ 2 = - \left (1-{{2M}\over{r^2}}\right )dt^2&+&\left (1-{{2M}\over{r^2}}\right )^{-1}dr^2 + \ r^2\ d\Omega^2 \ +\ 4M\ d\t^2\ \nonumber\\
&\pm& \ {{Q^2}\over{r^4}} \left (-dt^2 + dr^2 \right )\ .\label{KNewman-brane70}
\eea
Analysis reveals the significance of a two form over an one form in the formalism. It is in agreement with the fact that the non-linearity in a two form is larger than that in an one form, $i.e.\ M$$>$$Q$. With $a\neq 0$ the conserved quantities possess a lower bound and is described for $M$$\ge$$Q$$\ge$$a^2$ in the paper. It is believed that a space-time curvature was instantaneously generated by a two form on a pair of 
$(D{\bar D})_3$-brane creation at a Big Bang singularity \cite{spsk}. The charge $Q$ sourced by a lower form presumably couples to the curvature at a later time. The vacuum created brane and anti-brane in a pair moved in opposite direction.

\subsection{Emergent black holes: ${\mathbf{r^2>2M}}$ and ${\mathbf{r^2>Q}}$} 
In the regime the geometries on a non-BPS brane may seen to describe a quantum Kerr-Newman black hole in four dimensions.
For a calculational simplicity an emergent metric component $G_{rr}$ on a brane may further be fine tuned to: $r^4$$>$$>$$M^2$ and $r^8$$>$$>$$Q^4$. Apparently it truncates the ultra high energy modes in the emergent quantum patch $G_{rr}$ on a brane. Generically, the cutoff ignores the higher order iterative corrections in a perturbative gauge theory on a $D_3$-brane. However the cutoff does not affect the causal characteristics of the emergent quantum geometries sourced by a non-perturbative space-time curvature in the formalism. In fact the truncation in a metric component is consistent with the remaining components on a brane. In particular it owes its origin to five dimensions underlying a non-perturbative geometric construction on a $D_4$-brane. Thus a fine tuning may naturally be associated with a two form $U(1)$ gauge dynamics on a $D_4$-brane. 

\sp
\noindent
In the limit the $G_{rr}$ component in the brane geometries may be approximated to an appropriate form. Then the quantum vacua on a non-BPS brane, underlying a generalized curvature scalar on $S^1$ may explicitly be given by
\bea
ds^2&=&-\ \left (1-\frac{2M}{\rho_a^2} \pm  {{Q^2}\over{\rho_a^6}}\left [\tri_a-\frac{4r^2a^2\cos^2\t}{\rho_a^2}\right ]\right )dt^2\nonumber\\
&&+\ \left ( 1-{{2M-a^2\sin^2\t}\over{\rho_a^2}} \mp \frac{Q^2}{\rho_a^2\tri_a}\left [1-\frac{4r^2a^2\cos^2\t}{\rho_a^4}\right ]\right )^{-1}dr^2\nonumber\\
&&+\ \left (1+\frac{2M}{\rho_a^2}\left [ 1+{{\tri_a}\over{\rho_a^2}}\right ] + {{a^2\sin^2\t}\over{\rho_a^2}}\left [1- {{\tri_a}\over{\rho_a^2}}\right ]
\mp \frac{4a^2Q^2r^2\cos^2\t}{\rho_a^8}\right )\nonumber\\
&&\qquad\qquad\qquad\qquad\qquad\qquad\qquad\qquad\qquad\qquad \times \ \left (\rho_a^2\ d\t^2\right )\nonumber\\
&&+\ \left ( 1 \mp {{a^2Q^2}\over{\rho_a^6}}\left [\sin^2\t + \frac{4r^2a^2\cos^2\t}{\rho_a^2}\right ]\right )\tri_a\sin^2\t\ d\phi^2\nonumber\\
&&\pm\ \frac{2aQ^2}{\rho_a^6} \tri_a\sin^2\t dtd\phi \nonumber\\ 
&&-\ {{2\sqrt{2M}}\over{\rho_a}}\left (2M\left [ 1 + {{\tri_a}\over{\rho_a^2}}\right ] + \left [ 1 - {{\tri_a}\over{\rho_a^2}}\right ]a^2\sin^2\t\right )^{1/2}dtd\t \ .\label{KNewmann-brane7}
\eea
Two new emergent geometries differ primarily by a relative sign associated with $Q^2$-terms. They may seen to connect to each other through an electromagnetic duality. Metric causal patches with$+Q^2$ and $-Q^2$ respectively corresponds to an electric and a magnetic charged quantum geometries.
Analysis reveals that the emergent causal patches are associated with a relative wrong sign in the charge $Q$, when compared with a Reissner-Nordstrom black hole. The mixed quantum geometries may be separated out using a projection matrix ${\cal M}$ derived in ref.\cite{spsk,spsk-RN}. In the case we define the projection matrix as:
\be
{\cal M}=\frac{1}{2}\left( \begin{array}{ccc}
-{{G}_{tt}^+} && {{G}_{rr}^+}\\ && \\
{{G}_{rr}^-} && {-{G}_{tt}^-}
\end{array} \right)\ .\label{KNewman-brane8a} 
\ee
The inverse matrix takes a form: 
\be
{\cal M}^{-1}=\frac{1}{\det\cal M}\left( \begin{array}{ccc}
{{-G}_{tt}^-} && {{-G}_{rr}^+}\\ && \\
{{-G}_{rr}^-} && {{-G}_{tt}^+}
\end{array} \right)\ ,\label{KNewman-brane8}
\ee
\bea
{\rm where}\; {G}_{tt}^{\pm}&=&\left (1-\frac{2M}{\rho_a^2} \pm  {{Q^2}\over{\rho_a^6}}\left [\tri_a-\frac{4r^2a^2\cos^2\t}{\rho_a^2}\right ]\right )\nonumber\\
{\rm and}\; {G}_{rr}^{\pm}&=&\left ( 1-{{2M-a^2\sin^2\t}\over{\rho_a^2}} \pm \frac{Q^2}{\rho_a^2\tri_a}\left [1-\frac{4r^2a^2\cos^2\t}{\rho_a^4}\right ]\right )^{-1}.{}\label{KNewman-brane9}
\eea
In absence of a charge $Q$ the matrix determinant takes an explicit form:
$$\det {\cal M}= -{1\over{4\rho_a^2}}\left (8M - 2a^2\sin^2\t + a^4\sin^4\t \right )\ .$$
We compute the $\det {\cal M}$ at the Big Bang underlying a pair creation of a $D_3$-brane and an anti $D_3$-brane, with a nontrivial space-time curvature from a vacuum on a flat $D_4$-brane \cite{spsk}. The pair moved in opposite directions and are separated by a fifth dimension transverse to each other. The generalized curvature on a created pair of $(D{\bar D})_3$-brane underlie a geometric torsion and hence a rotation may seen to signal the beginning of a space-time on brane Universe. Since an event horizon coincides with an ergo radius at the poles the determinant at the beginning of a pair of brane and anti brane Universes simplify to yield:
$$\det{\cal M}=-\ {{2M}\over{\rho_a^2}}\ .$$ 
It may also be obtained in a limit $\rho_a^2\rightarrow \tri_a$ when  $Q= 0$ and $a\neq 0$. This in turn implies $G_{tt}=G_{rr}^{-1}$ and rules out the formation of an ergo sphere covering an event horizon. Alternately a nontrivial curvature arises, with a subtlety for $M\neq 0$, for the $S_2$-symmetry breaking parameter $a\neq 0$. Thus the matrix determinant is computed at the event horizon underlying a Big Bang to yield:
$$\det {\cal M}=\ -1$$ 
It re-assures a discrete transformation matrix at the creation of a pair of $(D{\bar D})_3$-brane Universe due to a two form fluctuations only. A charge  incorporates a quantum correction in addition to its dynamical feature into the quantum vacuum on a brane. In the next section 3.5 we shall see that a charge $Q$ does not modify the horizon of a black hole though it modifies an ergo radius or an angular momentum. The operation of the matrix on two independent column vectors may seen to project out the mixed emergent quantum geometries with lorentzian signatures on a $D_3$-brane. It may be given by
\be
{\cal M}\left( \begin{array}{c}
1\\
\\  
0
\end{array}\right)=\frac{1}{2}\left( \begin{array}{c}
{-{G}_{tt}^+}\\
\\
{{G}_{rr}^-}
\end{array}\right) 
\quad , \qquad {\cal M}\left( \begin{array}{c}
0\\
\\
1
\end{array}\right)=\frac{1}{2}\left( \begin{array}{c}
{{G}_{rr}^+}\\
\\
{-{G}_{tt}^-}
\end{array}\right )\ .\label{KNewman-brane-10}
\ee
The operation of the inverse matrix, on the same set of column vectors, projects out the required quantum patches from the mixed geometries on a brane. It may be given by
\be
{\cal M}^{-1}\left( \begin{array}{c}
1\\
\\  
0
\end{array}\right)=\left( \begin{array}{c}
{{G}_{tt}^-}\\
\\
{{G}_{rr}^-}
\end{array}\right) 
\quad , \qquad {\cal M}^{-1}\left( \begin{array}{c}
0\\
\\
1
\end{array}\right)=\left( \begin{array}{c}
{{G}_{rr}^+}\\
\\
{{G}_{tt}^+}
\end{array}\right )\ .\label{KNewman-brane11}
\ee
The inverse matrix projection corrects the relative sign of $Q^2$-term in the causal patches. It describes an euclidean black hole rotating with an imaginary angular momentum. Under an analytic continuation, the lorentzian quantum black hole is obtained with a real angular momentum. 

\subsection{Quantum Kerr(Newman) black hole in 4D}
Now we analyze the quantum geometries obtained via an inverse matrix projection containing the causal patches on an effective $D_3$-brane in type IIA/B superstring on $S^1$. A hint for fifth dimension transverse to a brane underlying the non-extremal black holes
implies the presence of an anti-brane. Thus the emergent black holes may be viewed through a global scenario in presence of a pair of  brane/anti-brane. A BPS brane and an anti BPS brane pair breaks the supersymmetry and leads to a non BPS brane configuration underlying the non-extremal geometries. We perform a transformation under an interchange $d\t \leftrightarrow d\phi$ into the ${\cal M}^{-1}$ projected geometries on a brane. The transformation primarily interchanges $\omega_{\t}\leftrightarrow \omega_{\phi}$. Then one of the brane geometries in the regime becomes
\bea
ds^2&=&\left (1-\frac{2M}{\rho_a^2} +  {{Q^2}\over{\rho_a^6}}\left [\tri_a-\frac{4r^2a^2\cos^2\t}{\rho_a^2}\right ]\right )dt_e^2\nonumber\\
&&+\ \left ( 1-{{2M-a^2\sin^2\t}\over{\rho_a^2}} + \frac{Q^2}{\rho_a^2\tri_a}\left [1-\frac{4r^2a^2\cos^2\t}{\rho_a^4}\right ]\right )^{-1}dr^2\nonumber\\
&&+\ \left ( 1 + {{a^2Q^2}\over{\rho_a^6}}\left [\sin^2\t + \frac{4r^2a^2\cos^2\t}{\rho_a^2}\right ]\right )\rho_a^2\ d\t^2\nonumber\\
&&-\ \frac{2aQ^2{\sqrt{\tri_a}}\sin\t }{\rho_a^5}\ dt_ed\t\nonumber\\
&&+\ \left (1+\frac{2M}{\rho_a^2}\left [1 +{{\tri_a}\over{\rho_a^2}}\right ] + {{a^2\sin^2\t}\over{\rho_a^2}}\left [ 1- {{\tri_a}\over{\rho_a^2}}\right ]
+ \frac{4a^2Q^2r^2\cos^2\t}{\rho_a^8}\right )\nonumber\\
&&\qquad\qquad\qquad\qquad\qquad\qquad\qquad\qquad\qquad\times\ \left ( \tri_a\sin^2\t\ d\phi^2\right )\nonumber\\ 
&&-\ \ {{2\sqrt{2M\tri_a}}\over{\rho_a^2}}\left (2M\left [ 1 + {{\tri_a}\over{\rho_a^2}}\right ] + 
\left [ 1 - {{\tri_a}\over{\rho_a^2}}\right ]a^2\sin^2\t \right )^{1/2}\nonumber\\
&&\qquad\qquad\qquad\qquad\qquad\qquad\qquad\qquad\qquad\ \times\ \left (\sin\t\ dt_ed\phi\right ).\label{KNewman-brane12}
\eea
The emergent charged geometries ensures $G_{tt}G_{rr}\neq 1$. It defines an ergosphere around the horizon(s). However the quantum geometry generically satisfies $G_{tt}G_{rr}=1$ at both the poles. It implies that a horizon and an ergosphere radius coincide at the poles. In a special case for $a$$=$$0$  the brane geometry satisfies $G_{tt}G_{rr}=1$. The ergo radius $r_{ergo}$ in the quantum Kerr black hole may be worked out from $G_{tt}=0$. The ergo radius may analytically be obtained from two independent conditions and they are given by
\be
\rho_a^4=2Q(r_{\rm ergo})\ a\cos\t \qquad {\rm and}\quad \rho_a^4 = {{Q^2}\over{2M}}\left ( r^2_{\rm ergo}+a^2\right )\ .\label{KNewman-brane121}
\ee
The quadratic equation in $r_{\rm ergo}$ is solved for its roots to yield:
\be
r^{\pm}_{\rm ergo}= {2aM_{\rm np}\cos \t} \left ( 1 \pm \sqrt{ 1 - {1\over{4{M_{\rm np}}^2\cos^2\t}}}\right )\ ,\label{KNewman-brane122}
\ee
where parameter $M$ is redefined to a non-perturbative mass $M_{\rm np}=M/Q$.
A lower bound on the parameters are set by $M$$>$$Q$$>$$a^2$. It shows that a small $Q$ underlying a linear charge describes a macroscopic black hole. A large $Q$ implies a non-linear charge and describes a microscopic black hole. ${M_{\rm np}}$ may be comparable with a typical black hole mass obtained in a dynamical metric theory. Thus, the background parameter $a$ in association with ${M_{\rm np}}$ defines a conserved angular momentum $J=a{M_{\rm np}}$ in the emergent charged black holes. The ergo radius further ensures a rotation to the charged black holes. Interestingly for a large mass $M$$>$$>$$Q$ the ergo radius in the quantum Kerr black hole (\ref{KNewman-brane12}) may be approximated to yield
\be
r^{+}_{ergo}\ \rightarrow\ 4J\cos\t - {{a^2}\over{4J\cos\t}}\qquad 
{\rm and}\quad r^-_{ergo}\ \rightarrow\ {{a^2}\over{4J\cos\t}}\ . \label{KNewman-brane123}
\ee
It shows that a charge $Q$ plays a significant role along with an intrinsic conserved quantity $M$ and defines a conserved angular momentum. 
A priori it leads to two different quantum rotating geometries on a non-BPS brane. Though $Q\neq 0$ in a non-perturbative curvature theory, a limit $Q\rightarrow 0$ may become sensible in a perturbative gauge theory. In the limit the outer ergo radius approaches infinity and the inner ergo radius disappears. Furthermore, the polar angle dependence of the ergo radii in the emergent charged black hole on a $D_3$-brane is analyzed at the poles to yield:
$$r^-_{ergo}={{a}\over{4{M_{\rm np}}}}\qquad {\rm and}\quad r^+=4a{M_{\rm np}}\left(1 -{1\over{16{M^2_{\rm np}}}}\right)\ .$$ 
The difference between the ergo radii decreases from their respective values at poles and they coincide $r^+_{ergo}\rightarrow r^-_{ergo} \rightarrow \infty$ on a equatorial plane. The shock wave peak in a limit $\t\rightarrow {\pi/2}$ ensures a quantum phase on a $D_3$-brane. The breakdown of the coordinate system in the limit presumably hints at an effective anti $D_3$-brane geometry which may be obtained from an effective $D_3$-brane under $r\rightarrow -r$. In other words the range of the spatial coordinates in the case ensures a global scenario with a pair of $(D{\bar D})_3$-branes. Thus the range of the polar coordinate on a $D_3$-brane is redefined with ($0<\t<\pi/2$) whereas the other half ($\pi/2<\t<\pi$) defines a ${\bar D}_3$-brane. 

\sp
\noindent
The other quantum geometry with $-Q^2$ in its causal patches (\ref{KNewmann-brane7}) under an interchange $d\t \leftrightarrow d\phi$ is described by
\bea
ds^2&=&\left (1-\frac{2M}{\rho_a^2} - {{Q^2}\over{\rho_a^6}}\left [\tri_a-\frac{4r^2a^2\cos^2\t}{\rho_a^2}\right ]\right )dt_e^2\nonumber\\
&&+\ \left ( 1-{{2M-a^2\sin^2\t}\over{\rho_a^2}} - \frac{Q^2}{\rho_a^2\tri_a}\left [1-\frac{4r^2a^2\cos^2\t}{\rho_a^4}\right ]\right )^{-1}dr^2\nonumber\\
&&+\ \left ( 1 - {{a^2Q^2}\over{\rho_a^6}}\left [\sin^2\t + \frac{4r^2a^2\cos^2\t}{\rho_a^2}\right ]\right )\rho_a^2\ d\t^2\nonumber\\
&&+\ \frac{2aQ^2{\sqrt{\tri_a}}\sin\t}{\rho_a^5}\ dt_ed\t \nonumber\\
&&+\ \left (1+\frac{2M}{\rho_a^2}\left [1 +{{\tri_a}\over{\rho_a^2}}\right ] + {{a^2\sin^2\t}\over{\rho_a^2}}\left [1 - {{\tri_a}\over{\rho_a^2}}\right ]
- \frac{4a^2Q^2r^2\cos^2\t}{\rho_a^8}\right )\nonumber\\
&&\qquad\qquad\qquad\qquad\qquad\qquad\qquad\qquad\qquad \times \ \left (\tri_a\sin^2\t\ d\phi^2\right )\nonumber\\ 
&&-\ {{2\sqrt{2M\tri_a}}\over{\rho_a^2}}\left (2M\left [ 1 + {{\tri_a}\over{\rho_a^2}}\right ] + a^2\sin^2\t 
\left [ 1 - {{\tri_a}\over{\rho_a^2}}\right ]\right )^{1/2}\nonumber\\
&&\qquad\qquad\qquad\qquad\qquad\qquad\qquad\qquad\qquad \times \ \left (\sin\t\ dt_ed\phi\right )\ .\label{KNewman-brane13}
\eea 
\begin{figure}
\mbox{
\subfigure{\includegraphics[width=0.38\textwidth,height=0.23\textheight]{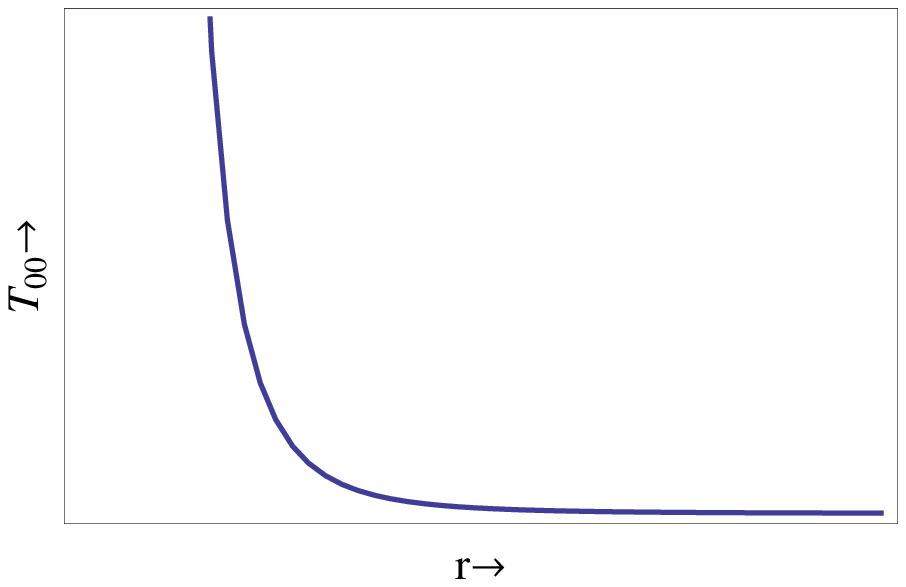}}
\hspace{.1in}
\subfigure{\includegraphics[width=0.5\textwidth,height=0.26\textheight]{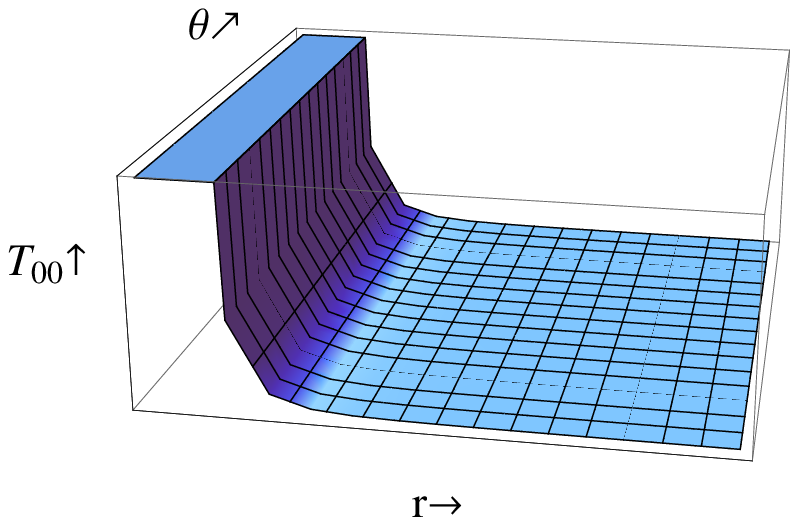}}}
\caption{\it The variations of $T_{00}$, respectively with $r$ for a fixed $\t$ and with $(r,\t)$ are shown for a quantum Kerr-Newman black hole in 4D.
The origin(s) is defined appropriately with $T_{00}=0$, $r=0$ and $\t=0$.}
\end{figure}

\noindent
We analyze the energy conditions for the 4D quantum black hole on a non-BPS brane. The energy momentum tensor, essentially sourced by the field components (\ref{KNewman-brane3}) is worked out in presence of a background Kerr(Newman) metric (\ref{KNewman-brane15}). The energy-momentum tensor consistent with a metric signature may formally be given by
\be
T_{\mu\nu}\ =\ {{-2}\over{\sqrt{-G}}} {{\delta S_{\cal F}}\over{\delta G^{\mu\nu}}}\ 
=\ \left ( {\cal F}_{\mu\lambda}{{\cal F}_{\nu}}^{\lambda} - {1\over{4}}G_{\mu\nu} {\cal F}^2\right )\ .\label{EC-1}
\ee
The $T_{00}$ component is explicitly computed for a Kerr(Newman) black hole. It is numerically analyzed for various ranges for $r$ on a given brane window and a variation of $T_{00}$ with $(r,\t)$ is explored in figure 1. Numerical analysis re-assures the energy condition $T_{00}\ge 0$ for a stringy Kerr(Newman) black hole.

\sp
\noindent
The ergo radii are worked out from two other independent conditions leading to $G_{tt}=0$. They are:
\be
\rho_a^6={{2Q^2r^2_{\rm ergo}}\over{M}}a^2 \cos^2\t\qquad {\rm and}\quad \rho_a^6= Q^2(r^2_{\rm ergo}+a^2)\ .\label{KNewman-brane131}
\ee
Apparently, the ergo radius turns out to yield:
\be
r_{\rm ergo}= a \left ( {{2a^2}\over{M}} \cos^2\t -1\right )^{-1/2}\ .\label{KNewman-brane132}
\ee
Then the angular velocity seems to be sourced by a background parameter. For $a^2$$>$$>$$M$ though the ergo radius $r_{\rm ergo}\rightarrow {\sqrt{{M}\over{2}}}$ at the poles, it becomes imaginary on the equator. However the emergent geometry (\ref{KNewman-brane13}) does not access the ergo sphere of a rotating black for $M$$\ge$$a^2$ on a $D_3$-brane. The brane geometries turn out to be defined with an euclidean time and hence possess an imaginary angular momentum. Nevertheless, a lorentzian signature may be re-established in the brane geometries with a real angular momentum under a subtle analytic continuation $t_e\rightarrow -it$. It may be noted that the time and space components in one form mixes within an ergo sphere just outside the event horizon of the black hole. The geometries for $M$$\ge$$a^2$ are worked out with a real angular velocity ($\Omega^{\t}$, $\Omega^{\phi}$) in lorentzian singnature. We obtain
\bea
ds^2&=&-\left (1-\frac{2M}{\rho_a^2} +  {{Q^2}\over{\rho_a^6}}\left [\tri_a-\frac{4r^2a^2\cos^2\t}{\rho_a^2}\right ]\right )dt^2\nonumber\\
&&+\left ( 1-{{2M-a^2\sin^2\t}\over{\rho_a^2}} + \frac{Q^2}{\rho_a^2\tri_a}\left [1-\frac{4r^2a^2\cos^2\t}{\rho_a^4}\right ]\right )^{-1}dr^2\nonumber\\
&&+\left (1+\frac{2M}{\rho_a^2}\left [1 +{{\tri_a}\over{\rho_a^2}}\right ] + {{a^2\sin^2\t}\over{\rho_a^2}}\left [ 1- {{\tri_a}\over{\rho_a^2}}\right ] +\frac{4a^2Q^2r^2\cos^2\t}{\rho_a^8}\right )\nonumber\\
&&\qquad\qquad\qquad\qquad\qquad\qquad\qquad\qquad\qquad \times \ \left (\tri_a\sin^2\t\ d\phi d\phi^+\right )\nonumber\\
&&+\left ( 1 + {{a^2Q^2}\over{\rho_a^6}}\left [\sin^2\t + \frac{4r^2a^2\cos^2\t}{\rho_a^2}\right ]\right )\ \rho_a^2\ d\t d\t^+\ .\label{KNewman-brane15}
\eea
and
\bea
ds^2&=&-\left (1-\frac{2M}{\rho_a^2} -  {{Q^2}\over{\rho_a^6}}\left [\tri_a-\frac{4r^2a^2\cos^2\t}{\rho_a^2}\right ]\right )dt^2\nonumber\\
&&+\left ( 1-{{2M-a^2\sin^2\t}\over{\rho_a^2}} - \frac{Q^2}{\rho_a^2\tri_a}\left [1-\frac{4r^2a^2\cos^2\t}{\rho_a^4}\right ]\right )^{-1}dr^2\nonumber\\
&&+\left (1+\frac{2M}{\rho_a^2}\left [1 +{{\tri_a}\over{\rho_a^2}}\right ] + {{a^2\sin^2\t}\over{\rho_a^2}}\left [ 1- {{\tri_a}\over{\rho_a^2}}\right ] -\frac{4a^2Q^2r^2\cos^2\t}{\rho_a^8}\right )\nonumber\\
&&\qquad\qquad\qquad\qquad\qquad\qquad\qquad\qquad\qquad \times \ \left (\tri_a\sin^2\t\ d\phi d\phi^+\right )\ \nonumber\\
&&+\left ( 1 - {{a^2Q^2}\over{\rho_a^6}}\left [\sin^2\t + \frac{4r^2a^2\cos^2\t}{\rho_a^2}\right ]\right )\ 
\rho_a^2\ d\t d\t^-\ ,\label{KNewman-brane16}
\eea
where $d\t^{\pm}= \left (d\t \pm \Omega^{\t} dt\right )$  and $d\phi^+=\left ( d\phi + \Omega^{\phi} dt\right )$. 
\begin{figure}
\mbox{
\subfigure{\includegraphics[width=0.38\textwidth,height=0.26\textheight]{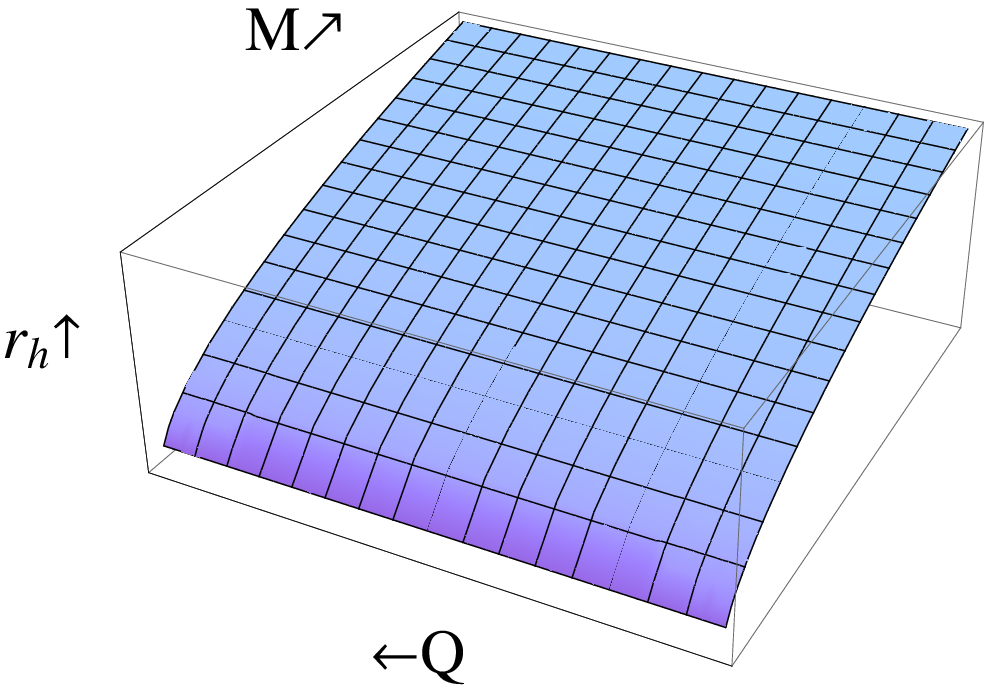}}
\hspace{.1in}
\subfigure{\includegraphics[width=0.5\textwidth,height=0.26\textheight]{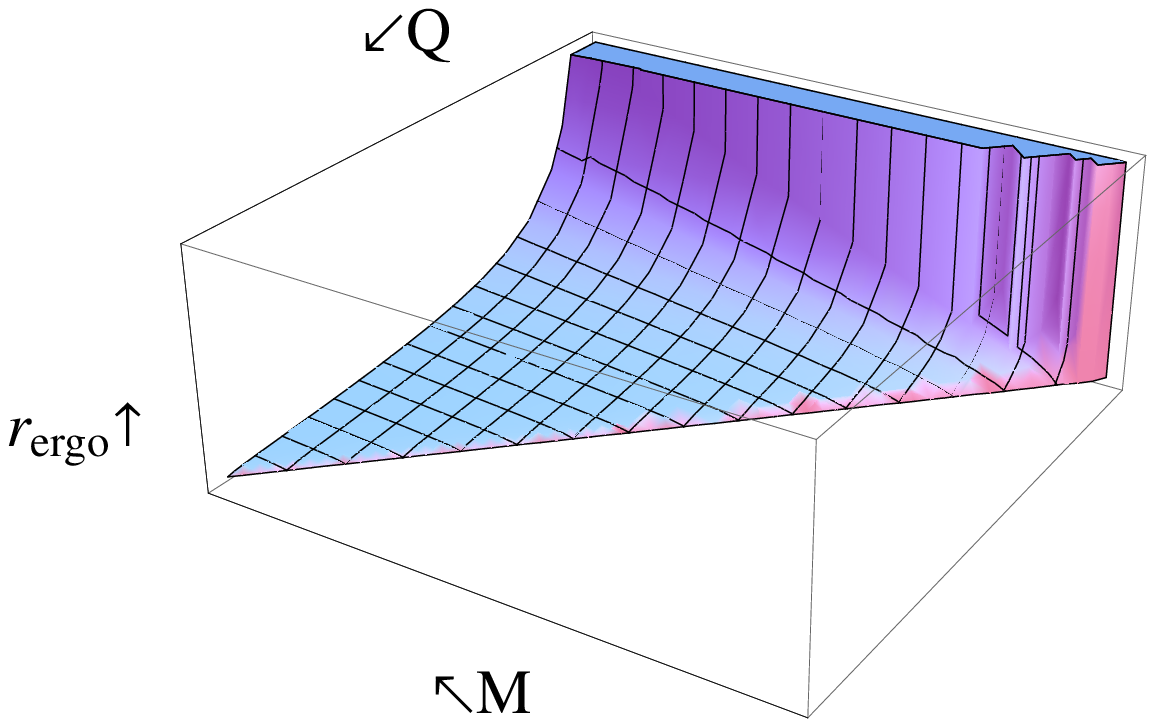}}}
\caption{\it The variations of an event horizon $r_h$ and an ergo sphere $r_{\rm ergo}$ in the moduli space of a quantum Kerr-Newman black hole are,  respectively, described by two distinct curved sheets in first and second figures. The event horizon and ergo sphere show an expansion with an increase in mass $M$ of a KN quantum black hole. Though the horizon is independent of a charge $Q$, the ergo sphere shrinks with an increased value or non-linearity of charge in the quantum Kerr vacuum.}
\end{figure}

\noindent
The  effective radial coordinate $\rho_a$ satisfies two independent horizon conditions. They are given by 
\be
\rho_a^2=2a\left ( r_h\cos\t_h\right )\qquad {\rm and}\quad \rho_a^2=\left ( 2M-a^2\sin^2\t_h\right )\ .\label{KNewman-brane17}
\ee 
Furthermore, these two conditions simplify at the horizon $r\rightarrow r_h$ to yield:
\be
a\cos\t_h \rightarrow r_h = \sqrt{2M-a^2}\ge 0 \qquad {\rm and}\qquad \tri_a^h\rightarrow 2M\ .\label{KNewman-brane171}
\ee
The charged rotating quantum black holes on a brane are generically characterized by an event horizon at $r\rightarrow r_h$$\ge$$0$ which implies  $\rho_a$$\rightarrow$$r_h{\sqrt{2}}$. Computation at the horizon shows a naive imaginary parameter: $(-ia)\sin\t_h\rightarrow {\sqrt{r_h^2-a^2}}$ associated with the off diagonal metric components in (\ref{KNewman-brane12}) and (\ref{KNewman-brane13}). For a minimal mass $M$, $i.e.\ 2M=a^2$, the imaginary parameter disappears and the event horizon may seen to be on an equatorial plane. An arbitrary mass $M$ satisfies $2M>a^2$ and hence the imaginary parameter turns out to be real in the lorentzian geometries (\ref{KNewman-brane15}) and (\ref{KNewman-brane16}). Strikingly the event horizon in a charged quantum black hole is independent of the charge $Q$. Thus the computations of physical quantity such as entropy at the black hole horizon remains unaffected by the charge. A charge independent horizon radius implies that its horizon area and hence the black hole entropy(S) does not depend on $Q$. A thermodynamic analysis under $dU=TdS$, reveals a constant internal energy for a large class of black holes defined with different $Q$. It leads to a degeneracy in a quantum Kerr-Newman geometries. 

\sp
\noindent
Alternately the quantum black hole may be viewed as an exact emergent geometry in a perturbative gauge theory. In fact both the quantum geometries are described by a common horizon which is independent of $Q$. However the ergo radii in a quantum black hole (\ref{KNewman-brane15}) are essentially defined by a charge. The ergo radius in (\ref{KNewman-brane16}) is independent of charge and is not accessible on a brane. It implies that the conserved charge may be identified with an angular momentum and is intrinsic to a non-perturbative geometric construction on a non BPS brane in the formalism. The conserved charge $Q$ may a priori seen to source an angular momentum which is defined by eq.(\ref{KNewman-brane18}). It is intrinsic to a non-perturbative geometric construction underlying an extra transverse dimension between a $D_3$-brane and an anti $D_3$-brane. A vacuum created pair of $(D{\bar D})_3$-brane possesses a space-time curvature and defines a non BPS configuration in the formalism. As has been argued, a non BPS brane nullifies the angular component $\Omega^{\t}$ and hence does not correspond to a new independent parameter in a global brane/anti-brane scenario. The characteristics of the quantum black holes presumably hint at a subtle interplay between $M$ and $Q$ underlying a generalized curvature ${\cal K}$ on $S^1$. They ensure an overlapping quantum geometry with: $r^-_{ergo}<r_h<r^+_{ergo}$. 

\subsubsection{Ring singularity and fluxes}
The $4D$ quantum Kerr-Newman geometries possesses a curvature singularity at $\rho_a\rightarrow 0$. It may also be obtained in a limit $r\rightarrow 0$ on the equator. A priori in the limit the non-trivial two form fields (\ref{KNewman-brane1})-(\ref{KNewman-brane2}) take a form:
\be
B_{tr}={{\sqrt{2M}}\over{a}}\; ,\qquad B_{r\t}=\left ( 4M -\left [2M + a^2\sin^2\t\right ]\sin^2\t \right )^{1/2}.\qquad {}\label{Knewman-brane172a}
\ee
\be
{\rm and}\qquad {\cal F}_{tr}= {Q\over{a^2\cos^2\t}}\ ,\qquad {\cal F}_{r\phi}= (a\ \sin^2\t)\ {\cal F}_{tr}\ .\qquad\qquad\qquad {} \label{KNewman-brane172}
\ee
On the equatorial plane the gauge curvature ${\cal F}_{\mu\nu}$ turns out to be singular in the limit $r\rightarrow 0$. In addition a background component $B_{r\t}\rightarrow r_h$ in the limit. Thus a curvature singularity in KN quantum vacua has its root in the fluxes. 
In absence of fluxes, $i.e.$ in a flat space-time (\ref{KNewman-brane4}), the curvature singularity is replaced by a coordinate ring singularity on a equatorial plane. In the limit $r\rightarrow 0$, all three circle equations underlying the 
KN quantum geometries imply a genuine ring singularity on the equator. For all other values of polar angle, $i.e.\ \t\neq \pi/2$, the radial coordinate $\rho_a$ may seen to be bounded from below by a non-zero minimal length for a real parameter $a$. Thus, a limit $\rho_a\rightarrow 0$ becomes unphysical when $\t\neq \pi/2$. In other words a KN quantum black hole possesses a genuine ring singularity only on a equatorial plane. It confirms the absence of a point singularity in the quantum Kerr black hole. A point singularity ($\rho_a\rightarrow 0$) in the quantum Kerr space-time may only be imagined for an imaginary $a$. The absence of a point singularity may provoke thought to imagine a plausible tunneling from a KN quantum black hole to a white whole vacuum via $\t\neq 0$ window.

\subsubsection{Killing vectors}
Now we compute the non-zero components of angular velocities in the KN quantum geometries at their respective horizon. They are:
\bea
\Omega^{\phi}&=&-2a {\sqrt{{2M-a^2}\over{M-a^2}}} \left ( {{\sqrt{2M^2+Ma^2- a^4}}\over{24M^2-12Ma^2-Q^2}}\right )\nonumber\\
&=& -\ 2ar_h^2{\sqrt{{2(r_h^2+3a^2)}\over{r_h^2-a^2}}} \times {1\over{\left [6r_h^2(r_h^2+a^2)-Q^2\right ]}}\qquad {}\nonumber\\
{\rm and}\qquad\quad &&\nonumber\\
\quad \Omega^{\t}&=&\pm Q^2{\sqrt{{2M(M-a^2)}\over{2M-a^2}}}\nonumber\\
&&\;\; \times\ {1\over{\left [ 8(2M-a^2)^3 + 2( 2M-a^2)a^2Q^2 - 2(M-a^2)Q^2 \right ]}}\nonumber\\
&=& \pm {Q^2\over{r_h}}{\sqrt{{r_h^4-a^4}\over{2}}} \times {1\over{\left [ 8r_h^6 + 2a^2Q^2r_h^2 - (r_h^2-a^2)Q^2 \right ]}}\ . \label{KNewman-brane18} 
\eea
The $(+)$-ve and $(-)$ve signs in $\Omega^{\t}$ respectively correspond to the Kerr geometries (\ref{KNewman-brane15}) and (\ref{KNewman-brane16}). 
The emergent black hole undergoes a rotation with an angular velocity $\Omega$ a priori defined with two non-zero components. It may be noted that  $\Omega^{\phi}$ is primarily sourced by the background parameter $a$, where as $\Omega^{\t}$ is sourced by a charge $Q$. An analysis in 
figure-1 infers that the charge $Q$ is highly non-linear and hence $\Omega^{\t}$ plays a significant role in addition to $\Omega^{\phi}$ in a quantum Kerr-Newman black hole in 4D. Under $r\rightarrow -r$ the angular components: $\Omega^{\t}\rightarrow -\Omega^{\t}$ and $\Omega^{\phi}$ remain unchanged and lead to a vanishingly $\Omega^{\t}$ in a global scenario underlying a pair of $(D{\bar D})_3$-brane. 
Under $a\rightarrow -a$ the $\Omega^{\phi}$ flips sign though the vacuum (\ref{KNewman-brane4}) on a $D_3$-brane remains invariant. It may imply that a pair of $(D{\bar D})_3$-brane underlying the nontrivial geometries are created from a flat vacuum with a Big Bang/Crunch \cite{spsk}. Thus both $\Omega^{\phi}$ and $\Omega^{\t}$ on a created pair of brane may seen to nullify in a global scenario. In other words $\Omega^{\phi}$ plays a significant role for $a\neq 0$ in a classical regime where as $\Omega^{\t}$ becomes significant for $Q\neq 0$ in a quantum geometry on a vacuumed created brane or an anti-brane within a pair.

\sp
\noindent
A fact that a pair of $(D{\bar D})_3$-brane, underlying non-trivial space-time curvature, has been created at the expense of a two form in an $U(1)$ gauge theory on a $D_4$-brane plays a vital role in the formalism. It implies that a two form on a $D_4$-brane sources a space-time on a pair of lower dimensional $D_3$-brane/anti-brane. Thus an effect of a two form is always defined via a global scenario with a pair of lower brane/anti-brane. The $\Omega^{\t}$ on a brane is annhiliated by an anti-brane. Then, the quantum Kerr(Newman) black hole on a pair of $(D{\bar D})_3$-brane is defined with a single angular component $\Omega^{\phi}$. The $t$ and $\phi$ isometries in the quantum Kerr(Newman) defines two Killing vectors and they are: $\xi_{(t)}={{\partial}\over{\partial t}}$ and  $\xi_{(\phi)}={{\partial}\over{\partial\phi}}$. A time-like Killing vector may be identified as: 
$K=\Big ({{\partial}\over{\partial t}} + \Omega^{\phi} {{\partial}\over{\partial\phi}}\Big )$.

\sp
\noindent
A priori the emergent quantum geometries (\ref{KNewman-brane15}) and (\ref{KNewman-brane16}) seem to differ in their direction in $\Omega^{\t}$ while they describe a  KN quantum black hole with a fifth transverse dimension sourced by an one form. Then a torsion geometry sourced by a five dimensional two form underlying the ansatz in (\ref{KNewman-brane1})-(\ref{KNewman-brane2}) may be expressed in terms of a large extra dimension and is formally given by
$$ds^2_{\rm 5D} =\ ds^2_{\rm KN}\ +\ L^2 d\psi^2\ .$$ 
Under an orthogonal rotation underlying a flip $\rho_a d\t\leftrightarrow L d\psi$ the emergent geometry on a vacuum created $D_3$-brane in presence of an extra dimension is analyzed to yield an additional Killing vector $\xi_{(\psi)}= {{\partial}\over{\partial\psi}}$. It implies a renaming of  $\Omega^{\t}\rightarrow \Omega^{\psi}$. In the case the time-like Killing vector turns out to yield:
$K_{(5D)}= \Big ({{\partial}\over{\partial t}} + \Omega^{\phi} {{\partial}\over{\partial\phi}} + \Omega^{\psi} {{\partial}\over{\partial\psi}}\Big )$. We define an orthogonal unit velocity vector $U^{\mu}\equiv \Big (\xi^{\mu }_{(t)} + \Omega^{\phi} \xi^{\mu}_{(\phi)} + \Omega^{\psi} {\xi^{\mu}_{(\psi)}}\Big )$ in five dimensions for a locally nonrotating observer with $U^2=-1$. We work out an orthogonality condition $G_{\mu\nu}U^{\mu}dx^{\nu}=0$ for a constant time surface. An absence of $G_{\phi\psi}$ component in the emergent brane geometry ensures the angular velocity components in eq(\ref{KNewman-brane18}). In section 5 we shall see that a quantum Kerr-Newman black hole in a low energy limit splits to describe two distinct semi-classical vacua, $i.e.$ a Kerr-Newman black hole and a Kerr black hole with a renormalized mass.

\subsubsection{Resolution of a conical singularity}
Naively the quantum geometries (\ref{KNewman-brane15}) and (\ref{KNewman-brane16}) may seen to possess a conical singularity due to a deficit polar angle $\delta_q$ in the formalism. For an instantaneous metric in (\ref{KNewman-brane15}) at an ergosphere, the deficit angle is worked out around $\t=0$ to order $\t^2$ by comparing with an appropriate line element:
$$ds^2=d\t^2 + \left (1-{{\delta_q}\over{2\pi}} \right )^2\t^2d\phi^2\ .$$ 
The deficit polar angle for a quantum Kerr black hole may be worked out to yield:
\be
\delta_q= 2\pi\left (1-\sqrt{{2\over{1+a^2}}\left (1 + {{Q^2}\over{a^2}}\left [16{\hat M}^2- 2{\hat M} +{{17}\over{16}} \right ]^{-1}\right )} \right )
\ .\label{CS-1}
\ee
Then the curvature scalar in the formalism may be expressed as:
$$ R_{\rm eff}= R(\rho_a)\ +\ 4\pi (1-n) \delta (\rho_a)\ , \qquad {\rm where}\quad n={{T_H}\over{T}}\ .$$
Needless to mention that a conical singularity becomes insignificant to a near horizon black hole ($T\rightarrow T_H$) underlying a non-linear $U(1)$ charge on a BPS $D$-brane. However on a non-BPS brane the conical singularity is generically associated with a curvature singularity which may be obtained in a limit $\rho_a\rightarrow 0$. For $\theta\neq \pi/2$ the ring (curvature) and conical singularities are not perceivable to an observer. The issue of apparent conical singularity on a equatorial plane at $r\rightarrow 0$ may seen to be resolved due to a non-contractible ring singularity there. 

\sp
\noindent
Generically a conical singularity does not arise in the formalism as the deficit angle (\ref{CS-1}) corresponds to a separation angle around the equatorial plane between a vacuum created pair of $D_3$-brane and and anti $D_3$-brane. This is in agreement with a fact that a brane and an anti-branes within a pair are respectively defined with a range of polar angle $0<\theta<\pi/2$ with $+r$ and $\pi/2<\theta<\pi$ with $-r$. Thus the deficit polar angle(s) leading to an apparent conical singularity on a equatorial plane becomes irrelevant in presence of an extra transverse dimension separating the vacuum created $D_3$-brane and ${\bar D}_3$-brane. 

\sp
\noindent
In fact a conical singularity does not depend on the space-time curvature. Thus in absence of charge $Q$, the deficit polar angle vanishes for $a^2=1$. It further re-assures the absence of a conical singularity in a flat vacuum underlying a non-degenerate quantum Kerr black hole. In other words, an apparent conical singularity may be an artifact of an extra dimension underlying a charge $Q$ in the formalism.

\subsection{Exact Kerr vacuum in 4D}
In this section we attempt to establish a relation between the two alternate and equivalent formulations on an effective $D_3$-brane. They are: (i) a non-perturbative quantum gravity theory defined by geometric torsion in a strong coupling regime and (ii) a two form U(1) gauge theory at weak coupling. Interestingly they establish a strong-weak coupling duality. A priori in a limit $Q\rightarrow 0$, the quantum black holes (\ref{KNewman-brane15}) and (\ref{KNewman-brane16}) drastically simplify to yield:
\bea
ds^2&=&-\left (1-\frac{2M}{\rho_a^2} \right )dt^2 +\left ( 1-{{2M-a^2\sin^2\t}\over{\rho_a^2}} \right )^{-1}dr^2 + \rho_a^2\ d\t^2\nonumber\\
&&+\left (1+\frac{2M}{\rho_a^2}\left [1 +{{\tri_a}\over{\rho_a^2}}\right ] + {{a^2\sin^2\t}\over{\rho_a^2}}\left [ 1- {{\tri_a}\over{\rho_a^2}}\right ]\right )\tri_a\sin^2\t\ d\phi^2\nonumber\\
&&-\ {{2\sqrt{2M\tri_a}}\over{\rho_a^2}}\left (2M\left [ 1 + {{\tri_a}\over{\rho_a^2}} \right ] + a^2\sin^2\t \left [ 1-{{\tri_a}\over{\rho_a^2}} 
\right ]\right )^{1/2}\nonumber\\
&&\qquad\qquad\qquad\qquad\qquad\qquad\qquad\qquad\qquad \times\ \left (\sin\t\ dtd\phi\right )\ .\label{KNewman-brane19}
\eea
The reduced geometry describes a quantum Kerr black hole with an event horizon at $r\rightarrow r_{h}$. The degeneracy in quantum Kerr-Newman black hole disappears in the limit.  The ergo sphere radius is computed to yield
\bea
r_{\rm ergo}&=&{\sqrt{2 M-a^2\cos^2\t}}\qquad {}\nonumber\\
&=&{\sqrt{r_h^2 +a^2\sin^2\t}}\ .\label{KNewman-brane192}
\eea
It shows that $r_{\rm ergo}\rightarrow r_h$ at the poles and $r_{\rm ergo}$$>$$r_h$ for $\t\neq0$. At the equator, $r_{\rm ergo}$ takes a maximum value which re-assures a rotation geometry intrinsic to the formalism. In the case, the angular velocity of the rotating black hole is precisely given by $\Omega^{\phi}$. The invariance of the emergent brane geometry under $r\rightarrow -r$ re-assures a global scenario underlying a $(D{\bar D})_3$ pair. 
The naive difference in the deformation geometry, along the $\phi$-coordinate, is due to the background fluctuations sourced by a two form. They do not contribute significantly to a propagating metric in an emergent scenario. It is interesting to note that a vanishing energy momentum tensor with the gauge choice for a two form in the gauge theory is in conformity with a vacuum solution in Einstein gravity.

\sp
\noindent
On the other hand an anti BPS brane and a BPS brane may seen to define a stable non-BPS brane in string theory \cite{sen1}. It is further re-confirmed by an extra dimension in the rotating quantum black hole. A $(D{\bar D})_3$-pair breaks the supersymmetry and describes a non-extremal black hole. The causal quantum geometric patches on a brane/anti-brane pair precisely identify with a typical Kerr black hole in Einstein gravity in presence of an extra dimension. The fifth dimension incorporates a local degree through a charge $Q$ to the background fluctuations in two form. Dimensional analysis confirms that a fifth dimension on $S^1$ of radius $R$ appropriately generates a gravitational constant $G^{(4)}$ in four dimensions from $G^{(5)}= G^{(4)}R$ in five dimensions. For a small scale, $i.e.\ r<<R$, the quantum black hole geometries (\ref{KNewman-brane15}) and (\ref{KNewman-brane16}) become significant. However for a large scale $r>>R$ the fifth dimension may be absorbed to define a four dimensional gravitational constant. Then, the emergent gravitational potential for $r>>R$ may be approximated to yield:
\be
{{2G^{(5)}M}\over{\rho_a^2}} \rightarrow {{2G^{(4)}MR}\over{\rho_a^2}}\approx  {{2G^{(4)}Mr}\over{\rho_a^2}}\ .\label{KNewman-brane193}
\ee
It hints at a scaling of $M$ in a semi-classical black hole (\ref{KNewman-brane19}) obtained in the large scale limit from a non-extremal quantum black hole. Under a scaling $M\rightarrow (Mr)$, the gauge choice (\ref{KNewman-brane1}) is modified without affecting the local degrees on a $D_3$-brane. Then the quantum geometry on a ${\bar D}_3$-brane with the modified ansatz may appropriately be obtained under $(r,M)\rightarrow -(r,M)$. Then a black hole on a non-BPS brane without any hidden dimension may be approximated to yield:
\bea
ds^2&=&-\left (1-\frac{2Mr}{\rho_a^2} \right )dt^2 +\left ( 1-{{2Mr-a^2\sin^2\t}\over{\rho_a^2}} \right )^{-1}dr^2 + \rho_a^2\ d\t^2\nonumber\\
&&+\left (1+\frac{2Mr}{\rho_a^2}\left [1 +{{\tri_a}\over{\rho_a^2}}\right ] + {{a^2\sin^2\t}\over{\rho_a^2}}\left [ 1- {{\tri_a}\over{\rho_a^2}}\right ]\right )\tri_a\sin^2\t\ d\phi^2\nonumber\\
&&-\ {2\sqrt{2{Mr\tri_a}}\over{\rho_a^2}}\left (2Mr\left [ 1+ {{\tri_a}\over{\rho_a^2}}\right ] + a^2\sin^2\t 
\left [ 1- {{\tri_a}\over{\rho_a^2}}\right ]\right )^{1/2}\nonumber\\
&&\qquad\qquad\qquad\qquad\qquad\qquad\qquad\qquad\qquad\; \times \ \left (\sin\t\ dtd\phi\right )\ .\label{KNewman-brane194}
\eea
The vanishing torsion in four dimensions under a gauge choice (\ref{KNewman-brane1}) reconfirms Riemannian curvature. In addition  a vanishing energy-momentum tensor ensures a vacuum geometry. Surprisingly the quantum geometry identifies with a typical Kerr black hole established as a vacuum solution in four dimensional Einstein gravity. Though it ensures an exact geometry the emergent metric components $G_{\phi\phi}$ and $G_{t\phi}$ differ from that in Kerr vacuum in Einstein gravity the difference may seen to modify the angular velocity computed at the event horizon. It is given by
\bea
\Omega^{\phi}_h&=&{{a}\over{3(r_h^2+a^2)}}\left (1 + {{4Mr_h}\over{r_h^2-a^2}}\right )^{1/2}\nonumber\\
&=&{{\Omega_{\rm Kerr}}\over{3}}\left ( 1 +\ 2\times {{r_h^2+a^2}\over{r_h^2-a^2}}\right )^{1/2}\ ,\label{KNewman-brane195}
\eea
where $\Omega_{\rm Kerr}$ signifies the angular velocity in a typical Kerr black hole in Einstein vacuum. The second term modifies the angular velocity and may be interpreted as a quantum correction. The modification is essentially due to a minimal length scale set by a two form and is intrinsic to the torsion dynamics on a non BPS brane. It may be checked that a $4D$ quantum Kerr black hole spins faster than its classical analogue for $6M<5a^2$ in the formalism. For a large black hole mass $i.e.\ 6M=5a^2$, the angular velocity precisely reduces to that in Einstein Kerr vacuum. Though the form of $\Omega_{\rm Kerr}$ is retained, it may also be expressed in terms of an intrinsic parameter ${\tilde M}$, $i.e.\ {\Omega_{\rm Kerr}}=a/(2Mr_h)$. Then, the angular momentum $J_{\rm Kerr}=a/(2r_h)$ in a four dimensional Einstein gravity receives a modification in the brane geometry.

\subsection{de Sitter to Kerr(Newman) universe} 
In this section we explore a plausible nature of space-time created on a pair of branes presumably at a Big Bang singularity. The space-time is believed to be created instantaneously with a minimal scale $2M=a^2$ in absence of $Q$ at the poles. Then the KN quantum geometries
(\ref{KNewman-brane12}) and (\ref{KNewman-brane13}) at the origin of a brane universe reduces drastically. A priori it is given by
\bea
ds^2=\left (1-\frac{a^2}{\tri_a} \right )dt_e^2 + \left ( 1-{{a^2}\over{\tri_a}}\right )^{-1}dr^2 &+& \tri_a d\Omega^2 + 2a^2\sin^2\t d\phi^2\nonumber\\
&-& 2a^2{\sqrt{2\over{\tri_a}}}\ \sin\t d\phi dt_e\ .\label{BigBang-1}
\eea
Filtering out the decoupled patches the quantum vacuum on a $D_3$-brane at the origin of Universe may be worked out for $r^2<a^2$ with $r^4<<a^4$. It  becomes 
\be
ds^2={{r^2}\over{a^2}}\left (1-\frac{r^2}{a^2} \right )dt_e^2+ {{a^2}\over{r^2}}\left ( 1-{{r^2}\over{a^2}}\right )^{-1}dr^2+ r^2 d\Omega^2
- {{\sqrt{2}}\over{a}}r^2\sin\t d\phi dt_e\ .\label{BigBang-2}
\ee
The geometry on an anti-brane may be obtained under $r\rightarrow -r$ and $a\rightarrow -a$. Then the space-time on a pair of $(D{\bar D})_3$-brane retains an $S_2$-symmetry and is given by
\be
ds^2= -\ {{r^2}\over{a^2}}\left (1-\frac{r^2}{a^2} \right )dt^2\ +\ {{a^2}\over{r^2}}\left ( 1-{{r^2}\over{a^2}}\right )^{-1}dr^2\ +\ r^2\ d\Omega^2
\ .\label{BigBang-3}
\ee
The Ricci scalar $R$ computed in the emergent Universe created on a pair of $(D{\bar D})_3$-brane is given by
\be
R= 2\left ( {1\over{r^2}} -{6\over{a^2}}+ {{15r^2}\over{a^4}}\right )\ .\label{BigBang-4}
\ee
In a limit $r\rightarrow a$ the curvature scalar $R= 20/a^2$ is a small positive constant for a cosmological scale `$a$'. The quantum Kerr(Newman) black hole in figure-1 indeed reassures a large scale for the parameter `$a$' in the microscopic description. The quantum vacuum (\ref{BigBang-3}) possesses a curvature singularity in a limit $r\rightarrow 0$. Interestingly the space-time at the creation of a brane-Universe is described by a four dimensional de Sitter vacuum in static coordinates. The $S_2$-breaking background parameter on a flat $D_3$-brane vacuum describes a cosmological horizon $r_c=a$. The limit $r\rightarrow 0$ may be identified with a Big Bang/Crunch. It may imply that a four dimensional de Sitter Universe was created, with a Big Bang from a flat non-spherical vacuum, by a two form in a non-linear $U(1)$ gauge theory. 

\sp
\noindent
The absence of fifth dimension at Big Bang further reassures a $(D{\bar{D}})$ pair creation there. Presumably the extra dimension grows in its length scale as a $D_{3}$ is moved away from a ${\bar{D}}_{3}$-brane in opposite direction. The length scale becomes large in TeV energy scale where quantum effects are still significant. A large extra dimension may helpful to understand laboratory quantum black hole in high energy experiments.
Analysis reveals that a quantum de Sitter undergoes tunneling among the degenerate quantum vacua. Furthermore an one form can vacuum create a pair of Kerr vacuum from a quantum Kerr. The process continues with an increase in electric energy until a critical electric field is reached where one finds a large degeneracy in a Kerr quantum vacuum. It is in conformity with a charge $Q$ independence of an event horizon in a Kerr(Newman) quantum vacuum discussed in section 3.4. In fact a black hole entropy $S$ formula ensures a $Q$ independence of entropy. Then a thermodynamic $dU=TdS$ analysis asserts a degeneracy in the vacua for the same internal energy at a constant temperature. In other words the degeneracies in a four dimensional quantum Kerr(Newman) are created at an expense of an electric field. It is indeed sourced by an extra dimension in a quantum Kerr vacuum which is intrinsic to the formalism. 

\section{Vacua from a nearly ${\mathbf{S_2}}$-symmetric brane} 
An extremely small background perturbation parameter $a$ to a spherically symmetric $S_2$-geometry on a $D_3$-brane may appropriately be defined for 
$a^2$$<$$<$$M$ with $a$$<$${\sqrt{M}}$. In the regime the quantum geometries on a pair of $(D{\bar D})_3$-brane shall be seen to reduce to 
Schwarzschild and RN quantum black holes respectively for a magnetic non-linear charge and a electric linear charge. Nevertheless the emergent quantum vacua are primarily sourced by a non-linear two form charge $M$ which has been identified with a mass of the quantum black holes.
The components of angular velocity computed in the limit of a very small $a$ for the emergent black holes (\ref{KNewman-brane15}) and (\ref{KNewman-brane16}) may drastically reduce to yield:
\be
\Omega^{\phi}\ \longrightarrow\ {{-4aM}\over{24M^2-Q^2}}\qquad {\rm and}\qquad \Omega^{\t}\ \longrightarrow\  {{\pm Q^2}\over{2{\sqrt{M}}(32M^2-Q^2)}}\ .\label{KNewman-brane20} 
\ee
It shows that the angular components $\Omega^{\phi}$ and $\Omega^{\t}$ are, respectively, sourced by a two form and an one form. The 
$\Omega^{\t}$ possesses its origin in a non-perturbative quantum correction which incorporates local degrees to the emergent geometry obtained in a gauge choice (\ref{KNewman-brane1}) and ({\ref{KNewman-brane2}). However in a non-perturbative generalized curvature theory on a non BPS brane, the $\Omega^{\t}\rightarrow 0$ which may also be reconfirmed from a redefinition of the black hole mass to absorb its charge $Q$. The flat vacuum (\ref{KNewman-brane4}) on a $D_3$-brane restores a spherical symmetry in a limit of a very small perturbation, $i.e.\ a^2$$<$$<$$M$. 

\sp
\noindent
The regime may seen to restore an $S_2$-symmetry on a flat $D_3$-brane and incorporates the significance of a charge $Q$ in the emergent quantum black holes. It may imply that a non-perturbative correction associated with $Q$ becomes vital to the black holes. In fact, an exact Kerr black hole (\ref{KNewman-brane19}) was shown to be equivalent to the quantum Kerr-Newman geometries (\ref{KNewman-brane15}) and (\ref{KNewman-brane16}) in presence of a non-perturbative correction. In other words, a restoration of $S_2$-symmetry in an exact Kerr black hole leads to the Schwarzschild and RN quantum black holes in presence of higher order fluxes.

\subsection{Quantum Schwarzschild black hole}
A priori a reduced quantum geometry is worked out for $a^2$$<$$<$$M$ and $a$$<$${\sqrt{M}}$ in eq(\ref{KNewman-brane16}) with $-Q^2$ term in the causal patches. The figure-2 assures a magnetic point charge $Q$ in a microscopic black hole which leads to a non-perturbative correction. However an one form linear charge may be gauged away in presence of a two form to describe a quantum Schwarzschild black hole. It may also be reassured by a fact that
a magnetic point charge does not incorporate any significant change in the characteristic properties of a quantum black hole. 

\sp
\noindent
In the context the higher order fluxes may seen to initiate a quantum tunneling of a RN to a large class of degenerate Schwarzschild vacua underlying a vacuum created pair of $(D{\bar D})_3$-brane. In particular the higher order fluxes may seen to modify the $G_{rr}$ component in the quantum regime. The quantum geometry (\ref{KNewman-brane22}), defined with a $-Q^2$-term in the causal patches may be worked out in presence of all high energy modes in $G_{rr}$ component. Ignoring a decoupled patch ($4M\sin^2\t$) in the $G_{\phi\phi}$ component from the quantum black hole we obtain
\bea
ds^2&=&-\left (1-\frac{2M}{r^2}\left [1 + {{Q^2}\over{2Mr^2}}\right ]\right )dt^2\nonumber\\ 
&&+\ \left ( 1-{{2M}\over{r^2}}\left [1+ \frac{Q^2}{2Mr^2}\right ]\right )^{-1}\left (1\ +\ {{\left (2Mr^2+Q^2\right )^2}\over{r^8}} + \dots\ \right )^{-1}dr^2\nonumber\\
&& +\ {{2\sin\t}\over{r}}\left ((2{\sqrt{2}})Md\phi - {{aQ^2}\over{r^3}}d\t\right ) dt + \ r^2 d\Omega^2\ .\label{KNewman-brane22}
\eea
The quantum geometry possesses an angular velocity defined by its two components. They are $\Omega^{\t}$ and $\Omega^{\phi}$ respectively sourced by a
charge $Q$ and mass $M$. It is characterized by an event horizon at
$$r\rightarrow r_e= {\sqrt{M}} \left [ 1 + {\sqrt{1+{{Q^2}\over{M^2}}}}\right ]^{1/2}=\ {\sqrt{\tilde M}}\ ,$$
where ${\tilde M}$ may be identified with a renormalized mass of a quantum black hole. In fact a 3D plot in figure-2 ensures that a microscopic black hole is indeed defined with a magnetic (linear) charge $Q$ which may be absorbed in a non-linear two form charge $M$ to define a renormalized mass $\tilde M$ for a Schwarzschild black hole underlying a fifth dimension. Interestingly the black holes mass computed for $M^4$$>$$>$$Q^4$ at the event horizon approximates to $${\tilde M}=2M\left ( 1 +{{Q^2}\over{4M^2}}\right )\ .$$ 
The renormalized mass may be approximated using a generic expression: 
\be
{\tilde M}=2M\left ( 1 +{{Q^2}\over{2Mr^2}}\right )_{r_e={\sqrt{\tilde M}}}\ .\label{KNewman-brane225}
\ee
It increases the minimal mass scale sourced by a two form in the formalism. Arguably, the $4D$ vacuum energy density ($M\rightarrow 0$) turns out to be a small positive constant presumably underlying a de Sitter brane sourced by a magnetic (point) charge. At this point we digress to recall that a two form is known to ensure a non-zero minimal length scale underlying a non-commutative parameter \cite{seiberg-witten}. In the regime the Schwarzschild quantum geometry on a $D_3$-brane may be re-expressed as:
\bea
ds^2&=&-\left (1-{{\tilde M}\over{r^2}}\right )dt^2\nonumber\\
&&+\ \left ( 1-{{\tilde M}\over{r^2}}\right )^{-1}\left (1\ +\ {{{\tilde M}^2}\over{r^4}} + \dots\ \right )^{-1}dr^2\nonumber\\
&&+\ {{2{\sqrt{2}}\sin\t}\over{r}}\left ({\tilde M}\ d\phi - {{Q^2}\over{r^2}}\ d\phi - {{aQ^2}\over{r^3{\sqrt{2}}}}\ d\t\right ) dt\ +\ r^2 d\Omega^2\ . \label{KNewman-brane223}
\eea
At a first sight a non-zero angular velocity in a quantum Schwarzschild black hole may appear to be counter-intuitive. In fact the angular velocity is sourced by the conserved charges $M$ and $Q$ in the formalism. Under $r\rightarrow -r$ and $a\rightarrow -a$ the emergent Schwarzschild black hole corresponds to that on an anti $D_3$-brane. Then the angular velocity on a brane is completely nullified by that on an anti-brane. In a global scenario  a Schwarzschild quantum black hole may be given by
\be
ds^2=-\left (1-{{\tilde M}\over{r^2}}\right )dt^2 + \left ( 1-{{\tilde M}\over{r^2}}\right )^{-1}\left (1 +{{{\tilde M}^2}\over{r^4}}+\dots\right )^{-1}dr^2+r^2 d\Omega^2\ . \label{KNewman-brane224}
\ee
The fifth dimensional hint in a four dimensional quantum Schwarzschild black hole is intrinsic to a generalized curvature underlying a propagating torsion in five dimensions. It further hints at the landscape vacua underlying a very large number of quantum black hole geometries on a pair of brane/anti-brane in superstring theory. In a low energy limit the higher orders from $1/r^4$ in a renormalized mass Schwarzschild may be ignored to obtain a typical Schwarzschild black hole sourced by a five dimensional potential in Einstein vacuum. It is in precise agreement with the notion of a mass in Einstein gravity underlying a superstring theory. A renormalized mass (\ref{KNewman-brane224}) in a Schwarzschild black hole ensures a local degree in the gauge choice underlying a generalized curvature theory in five dimensions. The correction to a brae mass $M$ becomes significant in a quantum regime and becomes small in a low energy regime. An emergent RN-black hole (\ref{KNewman-brane24}) and a Schwarzschild black hole 
(\ref{KNewman-brane224}) are analogous to that recently obtained by the authors \cite{spsk-RN} in presence of a set of different ansatz for the gauge fields. Analysis may imply that a vacuum created pair of $(D{\bar D})_3$-brane by a two form in a superstring theory may be a candidate 
to explore an emergent graviton on a $D_4$-brane. It further hints that the quantum gravity effects become significant in five and higher dimensions, in a geometric torsion theory.

\subsection{Quantum Reissner-Nordstrom black hole}
The remaining quantum geometry (\ref{KNewman-brane15}) defined with $+Q^2$ in the causal patches is described by electric charge in addition to a black hole mass $M$. The figure-2 may be revisited to conclude a non-linear $Q$. The emergent geometry may be worked out in presence of all the high energy fluxes in $G_{rr}$ components. Ignoring a decoupled patch the line-element may be given by
\bea
ds^2&=&-\left (1-\frac{2M}{r^2} + {{Q^2}\over{r^4}}\right )dt^2\nonumber\\
&&+\ \left ( 1-{{2M}\over{r^2}} + \frac{Q^2}{r^4}\right )^{-1}\left [1\ +\  {1\over{r^4}}\left (2M-\frac{Q^2}{r^2}\right )^2 + \dots\ \right ]^{-1}dr^2
\nonumber\\
&&+\ {{2\sin\t}\over{r}}\left ((2{\sqrt{2}})Md\phi + {{aQ^2}\over{r^3}}d\t\right ) dt\ +\ r^2 d\Omega^2\ . \label{KNewman-brane21}
\eea
The horizon radii are computed to yield: 
$$r_{\pm}= {\sqrt{M}} \left [ 1 \pm {\sqrt{1-{{Q^2}\over{M^2}}}}\right ]^{1/2}\ .$$
It hints at an extra dimension which is in conformity with a generalized curvature scalar 
${\cal K}$ on an effective $D_4$-brane. It describes a four dimensional RN-geometry. However it possesses an angular momentum unlike to that in a typical RN-black hole established in Einstein gravity. Nevertheless the angular momentum is intrinsic to a geometric torsion in five dimensions. A priori the angular velocity components computed at the event horizon in (\ref{KNewman-brane21}) is given by
\bea
&&\Omega^{\phi}={{-2M{\sqrt{2}}}\over{r_+(r^2_+ + 4M)a\sin\t_h}}\nonumber\\
{\rm and}&&\Omega^{\t}={{\pm 8Q^2}\over{({\sqrt{2M}})^5}} \left (1+ \sqrt{1-{{Q^2}\over{M^2}}}\right )^{-3}.\qquad {}\label{KNewman-brane23} 
\eea
Under $r\rightarrow -r$ and $a\rightarrow -a$  the emergent RN-geometry on an effective $D_3$-brane corresponds to that on an anti $D_3$-brane.
In a global scenario, $i.e.$ with a vacuum created pair of $D_3$-brane and an anti $D_3$-brane separated by an extra small dimension, $\Omega^{\phi}$ and $\Omega^{\t}$ on a brane nullifies that on the anti-brane. Thus the naive polar angle in $\Omega^{\phi}$ on a brane does not affect the quantum gravity on a non BPS brane. From the perspective of a higher dimensional theory it ascertains a need for $\t$-slicing geometry as $\Omega^{\phi}$ decreases from poles to the equator on a $D_3$-brane or a ${\bar D}_3$-brane. Then the quantum RN-geometry on a $(D{\bar D})_3$-brane may reduce to yield:
\bea
ds^2=&-&\left (1-\frac{2M}{r^2} + {{Q^2}\over{r^4}}\right )dt^2   \nonumber\\
&+& \left ( 1-{{2M}\over{r^2}} +\frac{Q^2}{r^4}\right )^{-1}\left (1+ {{f^2}\over{r^8}} + \dots\right )^{-1}dr^2\ +\ r^2 d\Omega^2\ .\label{KNewman-brane24}
\eea
where $f=\left (2Mr^2-Q^2\right )$. A priori the $S_2$-symmetry in the quantum RN-black hole may seen to be broken due to a deformation component ($4M\sin^2\t$) in $G_{\phi\phi}$ which is ignored in eq(\ref{KNewman-brane24}). Nevertheless a quantum RN-black hole retains the spherical symmetry
as the deformation component decouples from the rest of the geometry. In a low energy limit the non-linearity in $Q$ appears to decouple from a large mass black hole. In the limit a quantum RN-vacuum reduces to a semi-classical RN-black hole presumably corresponding to a vacuum in Einstein-Maxwell theory. In other words a four dimensional Einstein vacuum may seen to receive a non-perturbative correction in a generalized torsion theory on a $D_4$-brane. Arguably a non-perturbative correction possesses it origin in the non-linearity of the electric charge $Q$ sourced by a torsion in five dimensions.

\sp
\noindent
Apparently the phenomenon of Hawking radiation may be argued in a quantum RN-black hole to establish an equipotential between the sources characterized by the parameters ($M$ and $Q$). The radiation stops with an extremal RN-geometry for $Q\leftrightarrow M$. Ignoring the decoupled geometric component ($4Q\sin^2\t$) the remnant black hole described in an extremal limit takes the form of a Schwarzschild geometry. It is given by
\bea
ds^2&=&-\left (1-\frac{M}{r^2}\right )^2dt^2 + \left ( 1-{{M}\over{r^2}}\right )^{-2}\left (1 + {{4M^2}\over{r^4}}\left[1 -{{M}\over{r^2}}\right] + \dots\right )^{-1}dr^2\nonumber\\
&&\qquad\qquad\qquad\qquad\qquad\qquad\qquad\qquad\qquad\qquad\qquad +\ r^2 d\Omega^2\ .\label{KNewman-brane242}
\eea
In a low energy limit, defined by a large $r$, the extremal RN-black hole on a pair of $(D{\bar D})_3$-brane may reduce to yield:
\be
ds^2=-\left (1-\frac{M}{r^2}\right )^2dt^2\ +\ \left ( 1-{{M}\over{r^2}}\right )^{-2} dr^2\ +\ r^2 d\Omega^2\ .\label{KNewman-brane243}
\ee 
Analysis may suggest that a vacuum in Einstein gravity may emerge from the tunneling of vacua on a pair of $(D{\bar D})_3$-brane in a type II superstring theory.
\subsection{Quantum tunneling vacua}
We revisit a RN quantum black hole (\ref{KNewman-brane24}) to explore a renormalization scheme for its mass in an analogy with a Schwarzschild quantum black hole in section 4.1. A priori a renormalized mass ${\hat M}$ of a RN quantum black hole may formally be defined by
$${\hat M}= r_+^2= M \left [ 1 + {\sqrt{1-{{Q^2}\over{M^2}}}}\right ]\ ,\qquad {\rm where}\quad {{\hat M}}=2M\left ( 1 -{{Q^2}\over{4M^2}}\right )\ .$$ 
In fact the mass in a RN-black hole is computed at the outer horizon $r_+$ and may be approximated to yield:
\be
{\hat M}=2M\left ( 1 - {{Q^2}\over{2Mr^2}}\right )_{r_+={\sqrt{\hat M}}}.\label{KNewman-brane28}
\ee
Unlike in a quantum Schwarzschild (\ref{KNewman-brane225}) a renormalized mass ${\hat M}$ in a quantum Reissner-Nordstrom black appears to lower the vacuum energy in presence of an electric non-linear charge to a negative constant presumably leading to an AdS-brane in five dimensions. A renormalized $\hat M$ may as well be consistent with a propagating torsion in a $5D$ quantum Schwarzschild black hole whose causal patches may take a precise form in eq(\ref{KNewman-brane224}) for ${\tilde M}\rightarrow {\hat M}$. Thus an electric non-linear charge in a $4D$ quantum RN-black hole may be absorbed by a renormalized mass to describe a Schwarzschild quantum black hole in five dimensions underlying a geometric torsion dynamics. In other words a mass and a charge in a RN quantum black hole may be viewed via an unified Schwarzschild quantum black hole in $5D$.
The presence of a local degree in a torsion may alternately provide a clue to a quantum tunneling phenomenon between a RN- and Schwarzschild vacua in string theory. It may be noted that a torsion can unify an angular momentum and mass in a $4D$ RN quantum black hole to define a renormalized mass in a $5D$ Schwarzschild quantum black hole.

\sp
\noindent
Alternately the $4D$ quantum RN-black hole may further be worked out with an another renormalized mass described by the higher order corrections when compared with eq(\ref{KNewman-brane28}). In the case the renormalized mass may take a form:
\bea
M_{\rm r}&=&\left [{\hat M}\left ( 1 -{{{\hat M}}\over{r^2}} + {{{\hat M}^2}\over{r^4}} + \dots \right )\right ]_{r\rightarrow r_+={\sqrt{M_{\rm r}}}}\nonumber\\
&=&\left (2M -{{Q^2}\over{r^2}}\right )\ \times \nonumber\\
&&\left (\left[ 1 - {{2M}\over{r^2}}\right] + {{4M}\over{r^4}}\left[1 - {{Q^2}\over{r^2}}\right] + {{Q^4}\over{r^{4}}}\left[1 +{{Q^2}\over{r^{4}}}\right] + \dots \right )_{r\rightarrow {\sqrt{M_{\rm r}}}}\ .\label{KNewman-brane29}
\eea
Then a quantum RN-black hole (\ref{KNewman-brane24}) may be reviewed with a new event horizon expressed in terms of its high energy mass 
$M_{\rm r}$. It may be given by
\bea
ds^2&=&-\left (1-\frac{\hat M}{r^2}\right )dt^2\ +\ \left ( 1-{{{\hat M}}\over{r^2}} + {{{\hat M}^2}\over{r^4}}-{{{\hat M}^3}\over{r^6}}\right )^{-1}dr^2\ +\ r^2 d\Omega^2\nonumber\\
&=&-\left (1-\frac{\hat M}{r^2}\right )dt^2\ +\ \left ( 1-{{M_{\rm r}}\over{r^2}}\right )^{-1}dr^2\ +\ r^2 d\Omega^2\ .\label{KNewman-brane247}
\eea
It describes a $4D$ quantum Schwarzschild black hole with a renormalized mass $M_{\rm r}\le {\hat M}$ and an event horizon at $r_e={\sqrt{M_{\rm r}}}$. The presence higher order terms in $M_{\rm r}$ play a crucial role to invoke a quantum tunneling in a RN-black hole to reduce to a Schwarzschild vacuum.  An electric non-linear charge may seen to be manifested in the disguise of a spin in a quantum Schwarzschild black hole (\ref{KNewman-brane247}). It is further confirmed by the presence of an ergo sphere with $r_{\rm ergo}= {\sqrt{{\hat M}}}>r_e$. Unlike to a quantum Schwarzschild black hole (\ref{KNewman-brane224}) a tunneling leading to a new Schwazschild quantum vacuum (\ref{KNewman-brane247}) is a strictly a quantum phenomenon in four dimension. It can never lead to a classical Schwarzschild black hole in Einstein gravity.

\section{Semi-classical Kerr vacua in ${\mathbf{4D}}$}
A large $r$ limit in the KN quantum geometries (\ref{KNewman-brane15}) and (\ref{KNewman-brane16}) implies a low energy vacuum in the non-perturbative formalism. The quantum Kerr vacua truncate in the limit and may lead to Einstein Kerr vacuum in four dimensions. In particular,
a semi-classical limit is obtained from a KN quantum black hole in a lowest order approximation in $(1/r)$ when $M\neq 0$ and $Q\neq 0$. The truncation of all high energy geometric corrections in the limit shows a charge $Q$ dependence of horizon in the semi-classical black holes on a non-BPS brane. One of the large mass $M$ black holes may seen to be governed by a linear, charge $Q$ and leads to a typical KN black hole
in Einstein vacuum. The other may seen to be governed by a non-linear, charge $Q$ and leads to a Kerr black hole either in Einstein vacuum or in a non-linear gauge theory. The limit may be viewed through a large extra dimension between a pair of $(D{\bar D})_3$-brane.
\begin{figure}
\mbox{
\subfigure{\includegraphics[width=0.38\textwidth,height=0.2\textheight]{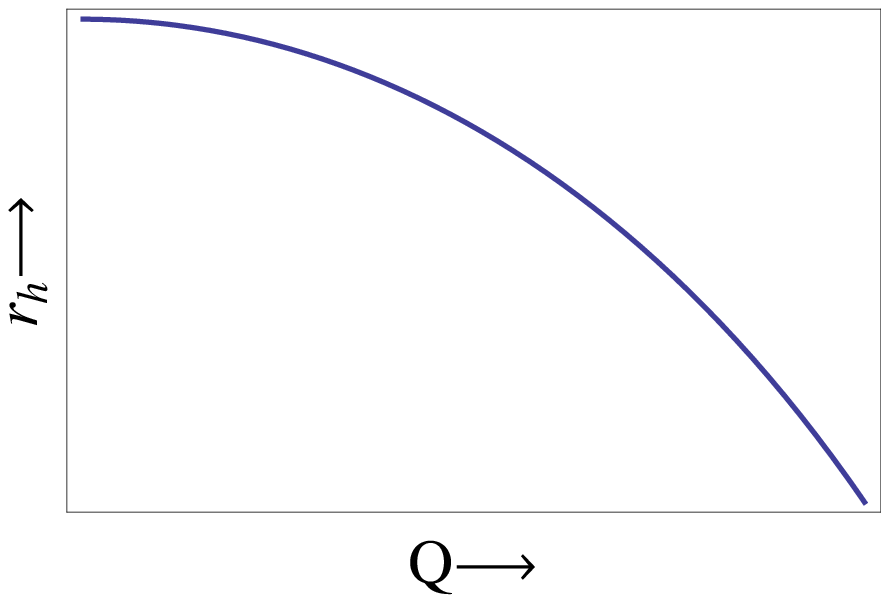}}
\hspace{.2in}
\subfigure{\includegraphics[width=0.5\textwidth,height=0.26\textheight]{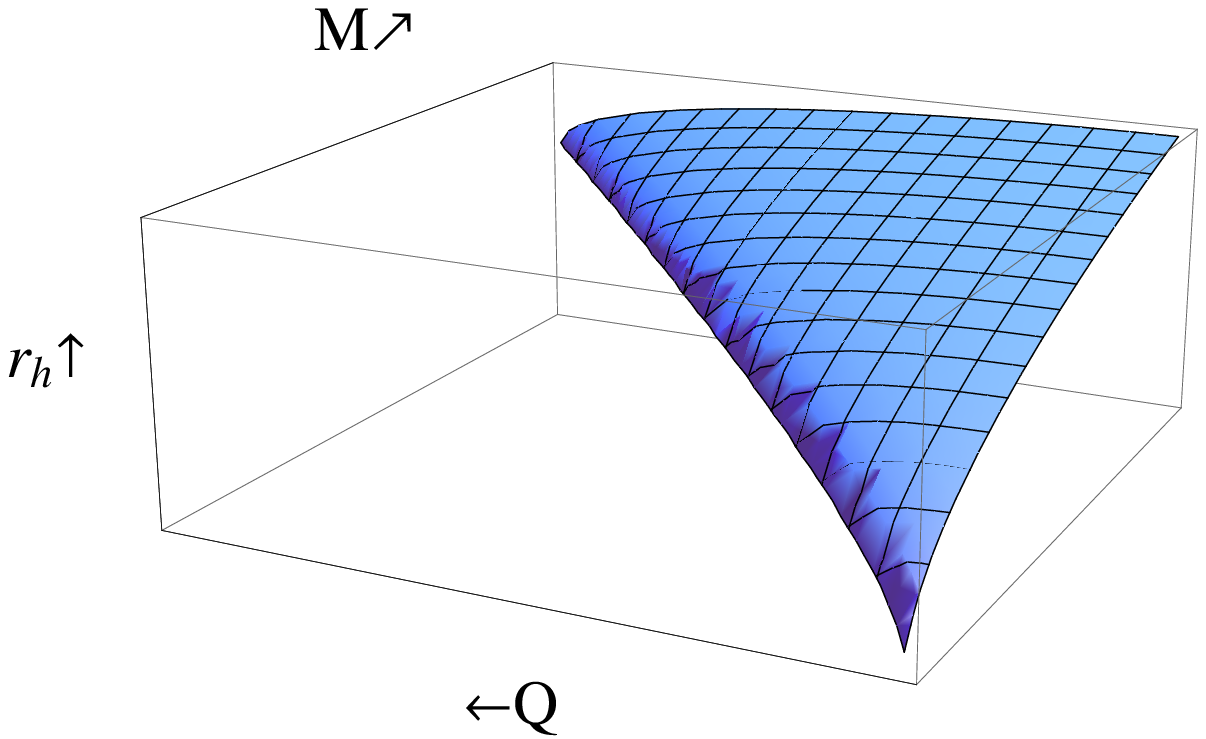}}}
\caption{\it A triangular folded sheet, underlying grid lines, shows an accessible physical horizon for various values in moduli space. 
The figures show that an expansion of a macroscopic KN black hole, obtained in a low energy limit from its quantum cousin,
is significantly larger for an increasing mass $M$ than for a change in value of its charge $Q$. The variation of its event horizon is identical to its ergo radius in the moduli space. It shows that a large mass rotating black hole is defined with a small or presumably a point charge in the semi-classical regime.}
\end{figure}
\subsection{Emergent Kerr-Newman black hole}
In a low energy limit the lowest order terms $(2M/\rho_a^2)$ and $(Q^2/\rho_a^4)$ become significant in the causal patches. In the limit the KN quantum geometry with $+Q^2$ term in eq.(\ref{KNewman-brane15}), reduces to yield:
\bea
ds^2&=&-\left (1-\frac{2M}{\rho_a^2} +  {{Q^2}\over{\rho_a^4}}\right )dt^2 + \left ( 1-{{2M-a^2\sin^2\t}\over{\rho_a^2}} + \frac{Q^2}{\rho_a^4}\right )^{-1}dr^2\nonumber\\
&&+\rho_a^2 d\t^2 + \left (1+\frac{4M}{\rho_a^2}\right )\tri_a\sin^2\t d\phi^2 -{{4M\sqrt{2\tri_a}}\over{\rho_a^2}}\sin\t dtd\phi\ .
\label{KNewman-brane30a}
\eea
The semi-classical geometry is characterized by two horizons at $r\rightarrow r_{\pm}$ and they are given by
\be
r_{\pm}= {\sqrt{{{r_h^2-a^2\cos^2\t_{\pm}}\over2}}} \left ( 1 \pm {\sqrt{1 + {{4a^2r_h^2\cos^2\t_{\pm} - 4Q^2}\over{\left (r_h^2-a^2\cos^2\t_{\pm}\right )^2}}}}\right )^{1/2}\ .\label{KNewman-brane34g}
\ee
It shows that an event horizon in a KN quantum black hole degenerates and develops a charge dependence in a low energy limit.
In particular the horizon radii on a equatorial plane is simplified to yield:
\be
r_+\ \rightarrow\ r_h\left ( 1 - {{Q^2}\over{2r_h^4}}\right )\qquad {\rm and}\quad r_-\ \rightarrow\ {{Q}\over{r_h}}\ .\label{KNewman-brane34}
\ee
Apparently the formal expression for the horizon radii may imply a shrinking KN quantum black hole in a low energy limit. However, the figure-3 for a macroscopic KN black hole on a pair of $(D{\bar D})_3$-brane ensures a linear electric charge $Q$ in four dimensions. 
As a result the KN quantum black hole may as well be described by a Einstein-Maxwell theory. The event horizon in a KN quantum black hole degenerates to describe an expanded event horizon and a shrinked inner horizon in a (semi)-classical limit. For a generic polar angle the event horizon retains its form (\ref{KNewman-brane34}) presumably with a redefined $r_h$ and $Q$. Furthermore in a limit $Q\rightarrow 0$ the degeneracy in horizon radius disappears and the semi-classical black hole reduces to a classical. It may a priori be identified with a quantum Kerr black hole in the generalized curvature theory. It is in conformity with our observation that the characteristics of a classical black hole may be identified with that in a Kerr quantum 
black hole. 

\sp
\noindent
In addition the semi-classical black hole (\ref{KNewman-brane30a}) is also described by two concentric ergo spheres. Their radii may be 
expressed in terms of the ergo radius defined in absence of a charge $Q$ in eq(\ref{KNewman-brane192}). The ergo radii are:
\be
r^{\pm}_{\rm ergo}= {\sqrt{{{r_{\rm ergo}^2-a^2\cos^2\t}\over2}}} \left ( 1 \pm {\sqrt{1 + {{4a^2r_{\rm ergo}^2\cos^2\t - 4Q^2}\over{\left (r_{\rm ergo}^2-a^2\cos^2\t\right )^2}}}}\right )^{1/2}\ .\label{KNewman-brane34g2}
\ee
For simplicity the ergo radii on a equatorial plane reduce to yield:
\be
r^+_{\rm ergo}\ \rightarrow\ r_{\rm ergo}\left ( 1 - {{Q^2}\over{2r_{\rm ergo}^4}}\right )\qquad {\rm and}\quad r^-_{\rm ergo}\ \rightarrow\ 
{Q\over{r_{\rm ergo}}}\ .\label{KNewman-brane342}
\ee
In general the ergo radii may be seen to retain the form (\ref{KNewman-brane342}) with a redefined charge $Q$ and $r_{\rm ergo}$. The redefined ergo radius in the case refers to a classical emergent Kerr geometry obtained in the limit $Q\rightarrow 0$ from a quantum Kerr black hole. Explicitly the redefined charge may be computed at the event horizon for $Q^2$$>$$>a^4$ with $Q$$>$$a^2$. It is approximated to yield:
\be
Q\ \rightarrow\  Q{\sqrt{1-{{2Ma^2}\over{Q^2}}}}\ .\label{KNewman-brane343}
\ee
Then a charge $Q$ modifies the ergo radii in a classical Kerr black hole (\ref{KNewman-brane192}). On the other hand, a charge
modifies the horizon radius in a quantum Kerr black hole. It ensures that the event horizon in the semi-classical black hole is completely bounded by the ergo spheres, $i.e.\ r^-_{\rm ergo}<r_-<r_+<r^+_{\rm ergo}$. The ergo sphere and horizon characteristics in a semi-classical black hole  are further supported by their causal geometric patches which identify to that in Kerr black hole on a equatorial plane. Thus a semi-classical rotating black hole on a non-BPS brane may correspond to a KN black hole established in Einstein vacuum. It reassures the generalized notion of fourth order curvature ${\tilde{\cal K}}_{\mu\nu\rho\lambda}$ established in the frame-work \cite{kpss1,kpss2,spsk,spsk-RN}. The $4D$ generic tensor ${\cal K}_{\mu\nu\rho\lambda}$ reduces to a Riemann tensor $R_{\mu\nu\lambda\rho}$ in a gauge choice for a two form leading to a non-propagating geometric torsion.

\subsection{Emergent Kerr black hole}
The KN quantum geometry (\ref{KNewman-brane16}) defined with $-Q^2$ term in its causal patches in a low energy limit simplifies to yield:
\bea
ds^2&=&-\left (1-\frac{2M}{\rho_a^2} -  {{Q^2}\over{\rho_a^4}}\right )dt^2 + \left ( 1-{{2M-a^2\sin^2\t}\over{\rho_a^2}} - \frac{Q^2}{\rho_a^4}\right )^{-1}dr^2\nonumber\\
&&+ \rho_a^2 d\t^2 + \left (1+\frac{4M}{\rho_a^2}\right )\tri_a\sin^2\t d\phi^2 - {{4M\sqrt{2\tri_a}}\over{\rho_a^2}}\sin\t dtd\phi\ . \label{KNewman-brane31}
\eea
A priori it describes a Kerr black hole with a non-linear magnetic charge. A lowest order approximation in $(Q^2/{\rho_a^4})$ incorporates the significance of charge $Q$ into the reduced semi-classical black holes. A higher order term $(Q^2/{\rho_a^8})$ leading to a quantum black hole unifies two distinct semi-classical black holes by removing relevance of $Q$ in a non-perturbative formalism. The semi-classical geometry (\ref{KNewman-brane31}) is characterized by a single horizon and an ergo sphere. The horizon radius is worked out to yield:
\be
r_{e}= {\sqrt{{{r_h^2-a^2\cos^2\t_+}\over2}}} \left ( 1 + {\sqrt{1 + {{4a^2r_h^2\cos^2\t_+ + 4Q^2}\over{\left (r_h^2-a^2\cos^2\t_+\right )^2}}}}\right )^{1/2}\ .\label{KNewman-brane34g5}
\ee
Though the event horizon in a quantum Kerr black hole develops a charge $Q$ dependence in a low energy limit, it does not seem to perturb the equivalence established between the independent potentials sourced by the form fields. It implies that a local degree in an one form are completely transferred into an otherwise non-dynamical two form in a non-linear gauge theory. On a equatorial plane the event horizon may be expressed in a simplified form:
\be
r_e\ \rightarrow\ r_h\left ( 1 + {{Q^2}\over{2r_h^4}}\right )
\ .\label{KNewman-brane345}
\ee
It reassures an expansion in the KN quantum black hole in a low energy limit. The ergo radius is given by
\be
r'_{\rm ergo}= {\sqrt{{{r_{\rm ergo}^2-a^2\cos^2\t}\over2}}} \left ( 1 \pm {\sqrt{1 + {{4a^2r_{\rm ergo}^2\cos^2\t + 4Q^2}\over{\left (r_{\rm ergo}^2-a^2\cos^2\t\right )^2}}}}\right )^{1/2}\ .\label{KNewman-brane34g6}
\ee
The ergo radius on a equatorial plane becomes:
\be
r'_{\rm ergo}\ \rightarrow\ r_{\rm ergo}\left ( 1 + {{Q^2}\over{2r_{\rm ergo}^4}}\right )\ .\label{KNewman-brane346}
\ee
\begin{figure}
\mbox{
\subfigure{\includegraphics[width=0.38\textwidth,height=0.23\textheight]{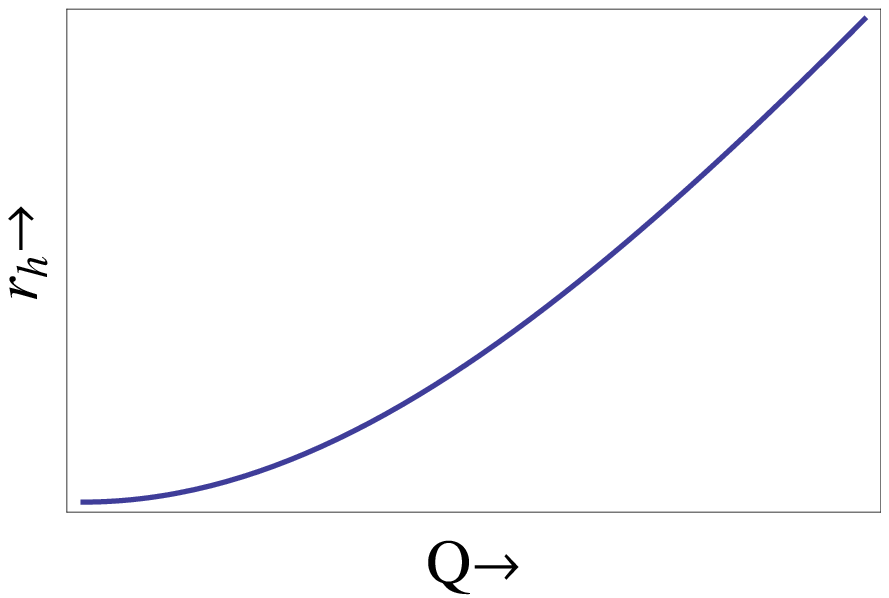}}
\hspace{.2in}
\subfigure{\includegraphics[width=0.5\textwidth,height=0.28\textheight]{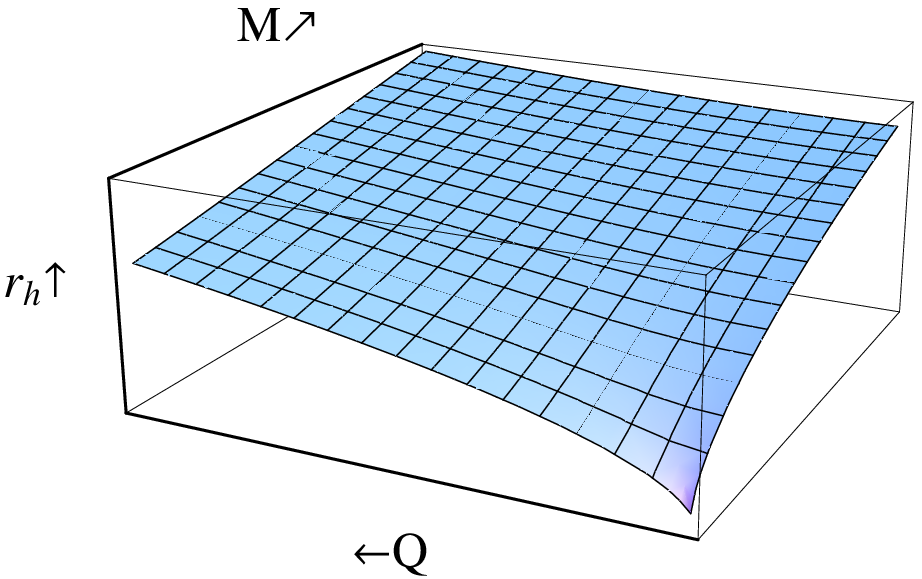}}}
\caption{\it The variations in event horizon and in ergo sphere show an identical behaviour in the moduli space a priori underlying a macroscopic charged Kerr black hole. An increase in horizon and ergo radii are reassured by a large or non-linear charge in addition to its large mass in the black hole. The mass is renormalized in presence of the charge or vice-versa to lead to a macroscopic Kerr black hole in Einstein vacuum or in a non-linear gauge theory.}
\end{figure}
In the case the event horizon is covered with an ergo sphere $i.e.\ r_+<r'_{\rm ergo}$. Analysis for a large mass $M$ black hole in figure-3 suggests that the event horizon and ergo radius increase with an increased value, or non-linearity, in charge $Q$. Unlike to a semi-classical Kerr-Newman black hole, a charged Kerr black hole (\ref{KNewman-brane31}) is a priori described by two non-linear charges ($M$ and $Q$) and a single horizon. It shows
that the potential due to a two form and that due to an one form are in equivalence at the horizon of the charged Kerr black hole. Thus one of the non-linear charges may be absorbed by the other without any change in the characteristics of the black hole. We appropriately renormalize the mass $M$ of a black hole to absorb a non-linear $Q$ in the causal patches. Then the charged Kerr geometry takes an explicit form and is given by
\bea
ds^2&=&-\left (1-\frac{2{\tilde M}_a}{\rho_a^2} \right )dt^2\ +\ \left ( 1-{{2{\tilde M}_a-a^2\sin^2\t}\over{\rho_a^2}}\right )^{-1}dr^2\nonumber\\
&&+\ \rho_a^2\ d\t^2\ +\  \tri_a\sin^2\t\ d\phi^2\nonumber\\
&&+\ {2\over{\rho_a^2}}\left (2{\tilde M}_a- {{Q^2}\over{2{\tilde M}_a-a^2\sin^2\t}}\right )
\left ({\sqrt{\tri_a}}\ \sin\t\ d\phi -{\sqrt{2}}\ dt\right )\nonumber\\
&&\qquad\qquad\qquad\qquad\qquad\qquad\qquad \times\ \left ({\sqrt{\tri_a}}\ \sin\t\ d\phi\right )\ ,\label{KNewman-brane315}
\eea
$${\rm where}\qquad {\tilde M}_a= M \left (1+ {{Q^2}\over{2M\rho_a^2}} \right )_{\rm horizon}\ .\qquad\qquad\qquad\qquad\qquad\qquad {}$$
\noindent
Explicitly, a bare mass may be expressed as:
\be
M= {\tilde M}_a- {{Q^2}\over{2\left (2\tilde M -a^2\sin^2\t_h\right )}}\ .\qquad {}\label{KNewman-brane316}
\ee
Thus at the expense of a magnetic non-linear charge $Q$ an emergent semi-classical Kerr black hole in a non-perturbative formulation describes a
dynamical gravity underlying Einstein vacuum. With a renormalized mass ${\tilde M}_a$ the characteristic features such as an event horizon, an ergo sphere and an angular velocity describe a Kerr vacuum in Einstein gravity. However the angular velocity in an emergent Kerr vacuum differs from that in Einstein vacuum. 

\sp
\noindent
Similarly we redefine a non-linear charge $Q$ to include mass $M$ in the charged Kerr geometry (\ref{KNewman-brane31}). 
The semi-classical emergent black hole may as well be interpreted in a non-linear gauge theory on a $D_3$-brane in the regime: $M$$>$$Q$$>$$a^2$. Qualitative analysis reassures a small value of the background parameter `$a$' for a large mass black hole. The low energy limit leads to 
${\tilde Q}^2$$>$$>$$a^4$ and the Kerr geometry in a non-linear gauge theory is given by 
\bea
ds^2&=&-\ \left (1-\frac{{\tilde Q}^2}{\rho_a^4} \right )dt^2\ +\ \left ( 1-{{{\tilde Q}^2}\over{\rho_a^4}} + {{a^2\sin^2\t}\over{\rho_a^2}}\right )^{-1}dr^2\nonumber\\
&&+\ \rho_a^2\ d\t^2\ +\ \tri_a\sin^2\t \ d\phi^2\nonumber\\ 
&&+\ \frac{4\delta{\tilde Q}^2 {\sqrt{2\tri_a}}\sin\t}{\rho_a^2\left (2{\tilde Q}- a^2\sin^2\t\right )}\ \left ( {\sqrt{\tri_a}}\ \sin\t \ d\phi - 
{\sqrt{2}}\ dt \right ) d\phi\ ,\label{KNewman-brane317}
\eea
\be
{\rm where}\qquad {\tilde Q}^2=Q^2 \left (1+ {{2M\rho_a^2}\over{Q^2}} \right )_{\rm horizon}.\qquad\qquad\qquad\qquad\qquad\qquad\qquad\qquad\qquad {}\label{KNewman-brane319}
\ee
The non-linear charges may be expressed as:
\bea
\delta{\tilde Q}^2&=&\left ({\tilde Q}^2-Q^2\right)\nonumber\\
&=&M\left (2{\tilde Q} - a^2\sin^2\t_h\right )\ .\qquad\qquad\qquad\qquad\qquad {}\label{KNewman-brane318}
\eea
The metric potential further hints at a fifth dimension. Analysis shows that a classical Kerr vacuum on a non-BPS brane possesses two equivalent descriptions. Their nontrivial space-time curvatures in eqs.(\ref{KNewman-brane315}) and (\ref{KNewman-brane317}) are, respectively, sourced by a renormalized mass ${\tilde M}$ in a non-perturbative curvature theory and a renormalized charge ${\tilde Q}$ in a non-linear $U(1)$ gauge theory. 

\subsection{Non-perturbative correction}
At the first sight it may appear to be interesting to compare the quantum Kerr-Newman geometries (\ref{KNewman-brane15}) and (\ref{KNewman-brane16}) with their respective vacua in low energy. Analysis may reveal the nature of a non-perturbative correction to an established Einstein vacua. A non-perturbative correction possesses its origin in an extra dimension between a vacuum created pair of $(D{\bar D})_3$-brane. 
In fact the gauge choice in (\ref{KNewman-brane1}) and (\ref{KNewman-brane2}) incorporates a local degree to a two form on a $D_4$-brane and ensures a dynamical geometric torsion in a five dimensional generalized curvature theory. Alternately, a two form in a four dimensional non-linear gauge theory creates a pair of brane underlying a non-trivial vacuum. An one form in the theory gives rise to a degeneracy in the quantum vacua. It may also be viewed through the respective non-perturbative corrections to the low energy vacua described by a Kerr-Newman black hole (\ref{KNewman-brane30a}) and a Kerr black hole (\ref{KNewman-brane31}). 

\sp
\noindent
The quantum Kerr-Newman (QKN) geometries (\ref{KNewman-brane15}) and (\ref{KNewman-brane16}) may be re-expressed in terms of their 
low energy black holes and their high energy deformation geometries. They are worked out in a regime and may formally be given by
\bea
ds^2_{\rm QKN}&=&ds^2_{\rm EG}\ +\ {{a^2\sin^2\t}\over{\rho_a^4}}\left ( 2M-a^2\sin^2\t\right )\nonumber\\
&&\times \Big [\tri_a\sin^2\t\ d\phi^2-{{1}\over{\sqrt{2}}} {\sqrt{\tri_a}}\ \sin\t\ d\phi dt_e + \cdots \Big ]\nonumber\\
&&\pm{{Q^2a^2}\over{\rho_a^6}} \Big [ \Big ( 1 -{{4r^2\cos^2\t}\over{\rho_a^2}}\Big )dt_e^2 +  
\Big ( 1 + {{4r^2\cos^2\t}\over{\tri_a}}\Big )dr^2\nonumber\\ 
&&+\Big (1 + {{4r^2a^2\cos^2\t}\over{\rho_a^2}}\Big )\rho_a^2 d\t^2 +
\Big ({{4r^2\cos^2\t}\over{\rho_a^2}}\Big )\tri_a\sin^2\t d\phi^2\Big ]\ ,\label{NP-1}
\eea
where $ds^2_{\rm EG}$ corresponds to a line-element in Einstein gravity. The $(+)$ve and $(-)$ve signs with $Q^2$ term in a non-perturbative correction respectively contribute to a semi-classical KN black hole (\ref{KNewman-brane30a}) and a classical Kerr black hole (\ref{KNewman-brane315}).
In addition to the classical black holes their $G_{\phi\phi}$ and $G_{t\phi}$ patches receive high energy deformations purely sourced by a two form in the formalism. They ensures a rapid growth in the spinning of the quantum black holes when compared with their low energy vacua. An absence of deformations to classical vacua along their causal patches reassure that the characteristic properties of a quantum black hole may be identified with its low energy vacuum. 

\sp
\noindent
Generically a non-perturbative correction in eq.(\ref{NP-1}) to Einstein vacuum underlying the quantum Kerr black holes is not flat. They may seen to share the curvature singularity at $\rho_a\rightarrow 0$ of the classical KN black holes. A priori a divergent quantum correction to Einstein vacua may not lead to a consistent quantum gravity description. However a resolution may seen to lie 
with a non-perturbative nature of correction arised in a quantum theory of gravity underlying an effective curvature scalar ${\cal K}$. We re-express the non-perturbative correction:
\bea
ds^2_{\rm NP}&=&\pm\ {{Q^2a^2}\over{\rho_a^6}}\left [-\ dt^2\ + dr^2\ + \rho_a^2\ d\t^2 \right ]\ \pm\ {{4Q^2a^2r^2\cos^2\t}\over{\rho_a^8}}\nonumber\\
&&\qquad \times\ \Big [ dt^2\ + {{\rho_a^2}\over{\tri_a}}dr^2\ +\ a^2\rho_a^2\ d\t^2\ +\ \tri_a\sin^2\t\ d\phi^2\Big ]
\ .\label{NP-2} 
\eea
The parameters in the Kerr(Newman) quantum black hole and its various limits are  
analyzed. The vacua may be summarized in a table and with a schematic diagram in figure 5.
\begin{table}
\begin{center}
\begin{tabular}{|c|cc|cc|}
\hline
Black holes on $(D{\bar D})_3$-brane, && Parameters:\\ 
{} && {$M>$ $Q>$ $a^2$} \\
\hline
Quantum Kerr(Newman) && Exact (degenerate) vacua:\\ 
&&small $M$, large $a$, independent of $Q$\\  \hline
Quantum Reissner-Nordstrom && Non-perturbative corrected vacuum\\
&& small $M$, $a\rightarrow 0$, non-linear $Q$\\ 
\hline
Quantum Schwarzschild && Non-perturbative corrected vacuum\\
&& small $M$, $a\rightarrow 0$, linear $Q$\\ 
\hline
(Low energy) Kerr-Newman && Einstein vacuum\\
&& large $M$, small $a$, linear $Q$\\ 
\hline
(Low energy) Kerr && large $M$, small $a$, non-linear $Q$\\  
\hline
Reissner-Nordstrom && large $M$, $a\rightarrow 0$, linear $Q$\\ 
\hline
Schwarzschild && large $M$, $a\rightarrow 0$, non-linear $Q$\\ 
\hline
\end{tabular}
\end{center}
\end{table}

\sp
\noindent
Interestingly a correction on the equatorial plane in a Kerr(Newman) black hole  may seen to be associated with a flat metric, which is in  
agreement with our other results \cite{spsk}-\cite{spsk-RN}. Since an observer comes across a ring singularity only on the equatorial plane of a $4D$ Kerr(Newman) black hole a non-perturbative correction becomes significant there. In other words a non-perturbative correction becomes sensible to a semi-classical vacuum corresponding to a macroscopic black hole. 

\begin{figure}
\begin{center}
\mbox{
\subfigure{\includegraphics[width=.65\textwidth,height=0.3\textheight]{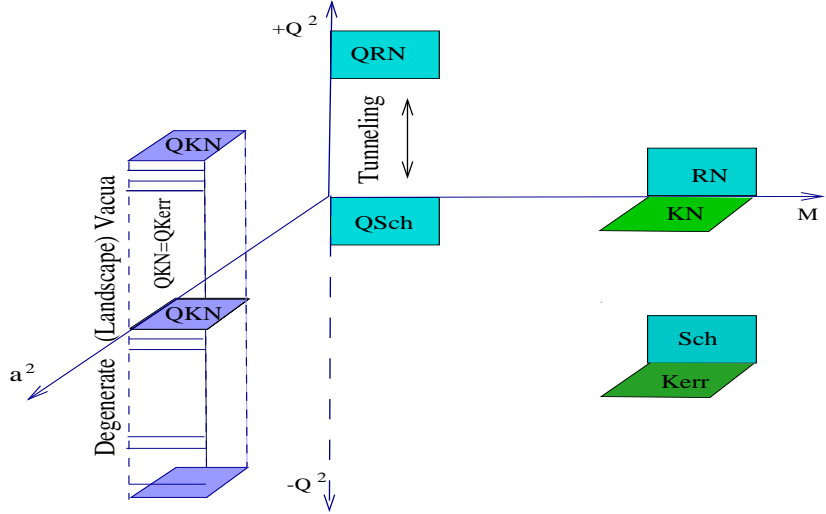}}}
\caption{\it A schematic diagram in moduli ($a^2, M, \pm Q^2$) space depicts quantum gravity vacua on a pair of $(D{\bar D})_3$-brane, separated by an 
extra dimension. The charge $Q$ independence of the horizon in QKN black hole make it equivalent to QKerr degenerate vacua in ($a^2,M$)-slices, presumably underlying a landscape scenario in superstring theory. The remaining two quantum vacua, QRN and QSch, in ($M,Q^2$)-slice are respectively defined for a non-linear $+Q^2$ and a linear $-Q^2$. A QRN is shown to tunnel to  QSch. Various classical vacua (KN, Kerr, RN and Sch) may be viewed in a low energy limit.}
\end{center}
\end{figure}

\sp
\noindent
On the other hand an independent nature of the correction to a space-time vacuum in Einstein gravity may allow one to view the quantum patches independently order by order in $1/\rho_a$ as obtained in eq.(\ref{NP-2}). The highest order term $1/\rho_a^8$ within a non-perturbative correction may be re-expressed under an interchange of its angular coordinates. As it has been argued in sections 3.2 and 3.4, a $d\theta\leftrightarrow d\phi$ interchange does not incorporate a change in characteristics of a quantum black hole. Then a geometric correction may be re-expressed as:
\bea
ds^2_{\rm NP}&=&\pm\ {{Q^2a^2}\over{\rho_a^6}}\left [-\ dt^2\ + dr^2\ + \rho_a^2\ d\t^2 \right ]\ \pm\ {{4Q^2a^2r^2\cos^2\t}\over{\rho_a^8}}
\nonumber\\
&&\qquad \times\ \Big [ dt^2\ + {{\rho_a^2}\over{\tri_a}}dr^2\ +\ \rho_a^2\ d\t^2\ +\ a^2\tri_a\sin^2\t\ d\phi^2\Big ]
\ .\label{NP-3} 
\eea
The line element associated with $Q^2/\rho_a^8$ turns out be flat in addition to the element with its lower order. Thus a generic non-perturbative correction to Einstein vacuum underlying the quantum effects in gravity may seen to be associated with a flat metric coupled to a non-linear charge sourced by an extra dimension in the formalism. A flat line-element in the quantum correction to a non-extremal KN black hole on a BPS brane is remarkable.

\section{Concluding remarks}
An effective spacetime curvature underlying a geometric torsion was revisited with a renewed interest to obtain a Kerr(Newman)  
quantum black hole in four dimensions on a vacuum created $D_3$-brane within a pair. A low energy perturbative string vacuum in presence of a non-perturbative $D$-brane correction is argued to define a vacuum created $D$-brane in the formalism. The scenario gave birth to an extra fifth dimension hidden between a vacuum created $D_3$-brane and an anti $D_3$-brane. It may provoke thought to believe that a two form may source an emergent graviton on a pair of $(D{\bar D})_3$-brane. On the other hand a geometric torsion on $S^1$ was shown to be sourced by a two form and an one form in the underlying $U(1)$ gauge theories respectively on a $D_3$-brane and an anti $D_3$-brane or vice-versa within a pair. The gauge fields are analyzed in a gauge choice to describe a quantum Kerr(Newman) black hole on a vacuum created pair of $(D{\bar D})_3$-brane at an event horizon. A gauge choice ensures a non-propagating torsion along with a propagating non-linear one form in the formalism. It may be interesting to check for a plausible strong-weak coupling duality underlying the two alternate descriptions leading to a torsion gravity theory and a non-linear gauge theory. The formalism may allow to realize a strong-weak coupling duality between a non-perturbative quantum gravity and a weakly coupled Einstein gravity. 

\sp
\noindent
One of the most interesting aspects of our analysis was described by the charge independence of an event horizon in the quantum Kerr(Newman) geometries. It was analyzed to argue for a degenerate Kerr vacua underlying a Kerr(Newman) quantum black hole in four dimensions. 
An equivalence between the Kerr(Newman) and Kerr quantum black holes was discussed to show the insignificance of a non-perturbative correction 
in the case. This in turn lead to an exact description underlying a degenerate Kerr vacua in the formalism. Nevertheless an electric non-linear charge $Q$, underlying a non-perturbative correction, has been shown to regain its significance in a limit leading to an extremely small $S_2$-symmetry breaking parameter `$a$' on a vacuum created $D_3$-brane. 

\sp
\noindent
The charge $Q$ was shown to couple to a flat metric and has been argued to incorporate a non-perturbative correction to a semi-classical black hole in Einstein gravity. A flat metric in a quantum correction to a semi-classical vacuum is remarkable. It implies that the quantum effects are 
associated with a non-linear charge sourced by an one form and they are independent of space-time curvature. A quantum correction was shown to 
play a vital role to incorporate local degrees into an emergent metric and makes its low energy vacua comparable to Einstein gravity. In fact a non-perturbative effects owe its origin to a vacuum created pair of $(D{\bar D})_3$-brane which underlie an extra fifth dimension. 
It plays a vital role in a gauge theory and presumably incorporates a non-perturbative quantum effects into a low energy perturbative string vacuum via a geometric torison in five dimensions. 

\sp
\noindent
A geometric torsion dynamics in presence of a flat metric may enlight us further with a perspective of quantum gravity. Intuitively the geometric construction in five dimensions underlying a non-perturbative quantum effect may provide clue to unfold the conjectured M-theory in eleven dimensions. Presumably an extra eleventh dimension between a vacuum created $D_9$-brane and an anti $D_9$-brane in type IIB superstring theory may help to explore the high energy domain in the non-perturbative M-theory. In addition a large extra spatial dimension could lower the quantum gravity scale and may lead to signal a ``graviton'' at LHC.

\sp
\noindent
In the limit of an extremely small background parameter `$a$' the degeneracy in the Kerr vacua disappears and they reduce to describe Schwarzschild and RN quantum black hole. It was shown that the mass of a charged quantum black hole is renormalized to absorb its charge $Q$. Thus a local degree is absorbed by a non-dynamical two form underlying a parameter $M$ in a gauge choice. It is believed to incorporate the dynamical aspects into the the gravitational potential in a Schwarzschild vacuum. The quantum geometry in a low energy limit was shown to describe a Schwarzschild vacuum in Einstein gravity. The other quantum black hole, described by two horizons, was worked out for a plausible mass renormalization computed at the event horizon to hint at a quantum tunneling of a RN black hole to a Schwarzschild vacuum. An absorption of an electric non-linear charge in a renormalized mass of a Schwarzschild black hole on a pair of $(D{\bar D})_3$-brane owes its origin in Poincare duality underlying a fifth dimension between the pair. In other words a non-dynamical two form, receives the local degrees from an one form within a gauge choice in four dimensions and, describes a propagating torsion in five dimensions. It was argued that a tunneling RN vacuum is purely a quantum phenomenon and has no analogue in a semi-classical RN in Einstein gravity.

\sp
\noindent
We have explored a low energy limit in the degenerate quantum Kerr(Newman) black hole. In the limit, an one form charge $Q$ has been shown to regain its significance at the expense of the degeneracy in the Kerr vacua. The quantum Kerr(Newman) deformations with $+Q^2$ reduce drastically to describe a 
semi-classical KN black hole presumably described by Einstein vacuum. A numerical analysis re-assures a linear (electric) charge $Q$ in an emergent macroscopic KN black hole underlying an effective curvature formalism. The other deformations with $-Q^2$ in the Kerr degenerate vacua 
in the limit have been shown to describe the low energy Kerr family in a geometric torsion theory as well as in a non-linear $U(1)$ gauge theory. An absorption of a non-linear magnetic charge in the bare mass makes it fat and re-assures a classical Kerr black hole. Needless to mention that the RN and Schwarzschild quantum black holes in the limit reduce to their counterparts in Einstein vacuum. 

\sp
\noindent
In the context it is note worthy to remark that the dark energy in our brane universe may have its source in a two form underlying a non-linear 
$U(1)$ gauge theory on a $D_4$-brane. Furthermore analysis suggests an equivalence between Einstein vacua to the low energy quantum vacua in a non-perturbative curvature theory on a non-BPS brane. The tunneling feature within the emergent quantum vacua on a brane hints at a tool to compute black hole entropy exactly. It may enhance our understanding on wall crossing formula relating a single centered black hole to a multi-centred one.

\section*{Acknowledgments}
S.S. acknowledges UGC and A.K.S. acknowledges CSIR for their fellowship. The work of S.K. is partially supported by a research grant-in-aid under the Department of Science and Technology, Government of India.

\def\anp{Ann. of Phys.}
\def\cmp{Comm.Math.Phys} {}
\def\prl{Phys.Rev.Lett.}
\def\jmp{J.Math.Phys.}
\def\prd#1{{Phys.Rev.}{\bf D#1}}
\def\jhep{JHEP\ {}}{}
\def\cqg#1{{Class.\& Quant.Grav.}}
\def\plb#1{{Phys. Lett.} {\bf B#1}}
\def\npb#1{{Nucl. Phys.} {\bf B#1}}
\def\mpl#1{{Mod. Phys. Lett} {\bf A#1}}
\def\ijmpa#1{{Int.J.Mod.Phys.}{\bf A#1}}
\def\mpla#1{{Mod.Phys.Lett.}{\bf A#1}}
\def\rmp#1{{Rev. Mod. Phys.} {\bf 68#1}}
\def\jaat{J.Astrophys.Aerosp.Technol.\ {}} {}


\end{document}